\documentclass[12pt]{article}

\usepackage{times}
\usepackage{physics}
\usepackage{graphicx}
\usepackage{amssymb}
\usepackage{amsthm}
\usepackage{amsmath}
\usepackage{dsfont}
\usepackage{bm}
\usepackage{mathrsfs}
\usepackage{bbold}
\usepackage{color}

\begin{document}

\title{Casimir cosmology}
\author{Ulf Leonhardt\\
Department of Physics of Complex Systems,\\
Weizmann Institute of Science,\\ Rehovot 7610001, Israel
}
\date{\today}
\maketitle

\begin{abstract}
In 1998 astronomers discovered that the expansion of the universe is accelerating. Somehow, something must have made gravity repulsive on cosmological scales. This something was called dark energy; it is described by Einstein's cosmological constant; and it amounts to about 70\% of the total mass of the universe. It has been conjectured that the cosmological constant is a form of vacuum energy, but its prediction from quantum field theory has failed by many orders of magnitude, until recently. Informed by empirical evidence on Casimir forces, Lifshitz theory has not only produced the correct order of magnitude, but is quantitatively consistent with the astronomical data. Moreover, the theory appears to resolve the tension between the measured and the predicted Hubble constant. There is therefore a good chance that Casimir physics explains dark energy. This article introduces cosmology for practitioners of vacuum forces as part of ``The State of the Quantum Vacuum: Casimir Physics in the 2020s'' edited by K. A. Milton. It may also be interesting for other physicists and engineers who wish to have a concise introduction to cosmology.
\end{abstract}

\newpage

\section{Introduction}

Observational cosmology has entered a golden age, the age of precision measurements. Gone are the days when ``cosmologists were often in error but seldom in doubt'' (Landau). Speculation and estimation have been replaced by precise data. What these data \cite{CMBPlanck} tell is astonishing: $95\%$ of the current content of the universe is completely unknown. Dubbed ``the dark sector'' (for want of a more illuminating term) it is described in the cosmological standard model, giving it its name $\Lambda$ Cold Dark Matter ($\Lambda$CDM) model. The true nature of the ``dark sector'' has been a mystery, or rather two mysteries, one less and the other more mysterious. For ``dark matter'' making up $25\%$ of the total mass density of the universe there are many theories and several experimental programmes to detect its particles \cite{LesHouches}. Yet the lion share of the ``dark sector'' --- $70\%$ of all mass --- has been an enigma \cite{LesHouchesLambda}. Called ``dark energy'' \cite{DarkEnergy}, it is described by Einstein's \cite{Einstein}  cosmological constant $\Lambda$ that causes a repulsive gravitational force. The constant $\Lambda$ is commonly believed to be associated with the quantum vacuum, its energy contribution is called the ``vacuum contribution'', but the actual mechanism of how the quantum vacuum creates the cosmological constant has not been clear \cite{WeinbergLambda} --- by quite a bit. Estimations from quantum field theory \cite{WeinbergLambda} are 120 orders of magnitude higher than the actual number \cite{CMBPlanck}. But these are based on a naive picture of vacuum fluctuations. Thanks to experimental and theoretical research on the Casimir effect \cite{Milonni,Milton,KMM,BKMM,Rodriguez,Lambrecht,Buhmann,Forces} we know a great deal more now about vacuum fluctuations. It is time for the practitioners of vacuum forces, the Casimir community, to take up the torch and shed light on ``dark energy''. 

The first steps have been made \cite{Annals,London,Berechya} and they seem encouraging. The 120 orders of magnitude in the mismatch between theory and measurement  disappear if one accepts the idea that the quantum vacuum needs to be renormalized. Renormalization is normal in practical Casimir physics \cite{Milonni,Milton,KMM,BKMM,Rodriguez,Lambrecht,Buhmann,Forces} and its absence would contradict experimental facts \cite{London}. What remains after renormalization in cosmology \cite{Annals} is not exactly zero, as simple calculations would tell \cite{Annals,Volovik,Ryskin} and also not approximately zero, as experience would suggest --- the Casimir force typically acts on the nanoscale, so how can it play any role on cosmological scales? The theoretical $\Lambda$ \cite{Annals} is of exactly the right order of magnitude. Moreover, the theoretical prediction is consistent with the astronomical data up to the level of precision of that data \cite{Berechya}. Furthermore, it fits and explains a tension in the astronomical data (and their interpretation) known as the Hubble tension \cite{DiValentino} that with each refinement in data and data analysis  just gets sharper and has recently reached crisis level \cite{Riess22}. The astrophysics community has discovered this and other inconsistencies between the $\Lambda$CDM model and the data, that urgently need explanations. The Hubble tension \cite{DiValentino} has been called the biggest crisis in contemporary astrophysics. But every crisis is an opportunity, and if the Casimir effect does indeed explain the cosmological constant, this is the golden opportunity for the Casimir community to make a decisive contribution to astrophysics.

How the Casimir effect may play a role in cosmology was explained elsewhere \cite{London}. Here I focus on the cosmology. This is primarily for bringing Casimir practitioners up to speed in a field that is normally alien to them, and secondly, for working out where exactly in the cosmic expansion the Casimir effect enters. There are excellent textbooks on cosmology, see {\it e.g.}\ Refs.\ \cite{Peebles,Harrison,Peacock,Mukhanov,Weinberg,Durrer,Dodelson}, but it takes time and effort to extract and absorb the parts relevant to Casimir physics. Here is hopefully a primer that is sufficiently concise and still sufficiently correct. Following Einstein's well--known advice, I have tried my best to write it ``as simple as possible, but not simpler''.

\section{Principles of cosmology}

Averaged over cosmological scales $ \gtrsim 100\mathrm{Mpc}$ space is homogeneous and isotropic. This is an empirically known fact (see e.g. Ref.~\cite{Gott}) that previously was theoretically postulated as the cosmological principle \cite{LL2}. The pc (parsec) is an astronomical unit of distance with $1\mathrm{pc}=3.26 \mathrm{ly}$ (light--years). To set the parsec into perspective, the nearest stars are a few pc away and galaxies like ours are in the order of $10\mathrm{Kpc}$ large. It is also known empirically \cite{CMBPlanck} that on cosmological scales space is flat to an excellent approximation. But the universe expands, the measure of length $\mathrm{d}\ell$ grows with time $t$ as some $\mathrm{d}\ell_0$ multiplied by the dimensionless scale factor $a(t)$. We condense these facts in the space--time metric:
%%%%%%
\begin{equation}
\mathrm{d}s^2 = c^2 \mathrm{d}t^2 - a^2 \mathrm{d}\bm{r}^2 \,,\quad
\mathrm{d}\bm{r}^2 =\mathrm{d}x^2+\mathrm{d}y^2+\mathrm{d}z^2 \,,\quad
a=a(t) \,.
\label{eq:metric}
\end{equation}
%%%%%%
The metric (\ref{eq:metric}) is called the Friedman--Lemaitre--Robertson--Walker metric \cite{Friedman1,Friedman2,Lemaitre1,Lemaitre2,Robertson1,Robertson2,Walker}. As usual, $c$ denotes the speed of light in vacuum. Any space--time geometry is a geometry of time: according to general relativity \cite{LL2} a metric $\mathrm{d}s$ describes the increment $\mathrm{d}s/c$ of the proper time of an object moving with spatial increment $\mathrm{d}\bm{r}$ per coordinate time $\mathrm{d}t$ \cite{LL2}. In our case, the time coordinate $t$ is called the cosmological time and the three--dimensional coordinates $\bm{r}$ are called the comoving coordinates. In astronomy the present time is denoted by $t_0$ and the scale factor is set to $a(t_0)=1$. Note that the metric takes the form of Eq.~(\ref{eq:metric}) only in frames comoving with the universe; the universe as a whole distinguishes a preferred reference frame. This does not mean that the laws of nature are frame--dependent, but rather that the initial conditions were such that space is uniform and flat on cosmological scales in some frame. In astrophysics, this frame manifests itself in the Cosmic Microwave Background (CMB) --- more details on the CMB in Sec.~4 and Appendix A.

Cosmological time is the time $\mathrm{d}s/c$ measured by an observer at rest ($\mathrm{d}\bm{r}=0$) with the CMB. We can define a new time coordinate t$\tau$ as
%%%%%%
\begin{equation}
\tau = \int \frac{\mathrm{d}t}{a}
\label{eq:tau}
\end{equation}
%%%%%%
(often denoted by $\eta$ in the literature) in which the metric appears as 
%%%%%%
\begin{equation}
\mathrm{d}s^2 = a^2(c^2 \mathrm{d}\tau^2 - \mathrm{d}\bm{r}^2) \,.
\label{eq:conformal}
\end{equation}
%%%%%%
The prefactor $a^2$ serves as a conformal factor in the metric --- the space--time of the expanding universe is conformally flat \cite{LeoPhil} --- and so $\tau$ is called conformal time.  As light rays propagate with $c=\mathrm{d}\ell/\mathrm{d}t$ (and $\mathrm{d}\ell=a\,\mathrm{d}r$) they follow lines of zero proper time ($\mathrm{d}s=0$). In this case the conformal factor is irrelevant: in conformal time $\tau$ and comoving spatial coordinates $\bm{r}$ light propagates like in empty, non-expanding space. This feature remains true beyond the approximation of geometrical optics and applies to full electromagnetic fields and their quantum fluctuations, as Maxwell's equations are conformally invariant \cite{LeoPhil}.

Although light propagates like in non--expanding space $\bm{r}$ with respect to conformal time $\tau$, this does not mean that the frequency of light observed from a source in the past is the same as the emitted frequency, because these frequencies are frequencies with respect to cosmological time. They are given by $\omega=-\partial_t\varphi$ where $\varphi$ denotes the phase. What is the same  in past and present is the frequency with respect to conformal time, $-\partial_\tau\varphi$. From $\partial_\tau = a\partial_t$ follows that the product $a\omega$ remains constant. In the expanding universe, the observed frequency $\omega$ gets lower and the wavelength $\lambda$ larger with growing $a$, which is known as the cosmological redshift. The redshift is abbreviated by $z$ and defined according to
%%%%%%
\begin{equation}
1+z = \frac{\lambda_\mathrm{obs}}{\lambda_\mathrm{emit}} = \frac{\omega_\mathrm{emit}}{\omega_\mathrm{obs}} \,,
\end{equation}
%%%%%%
which gives [for $a(t_0)=1]$:
%%%%%%
\begin{equation}
z=a^{-1}-1 \,,\quad a = \frac{1}{1+z} \,.
\label{eq:z}
\end{equation}
%%%%%%
The redshift is measured by comparing the observed spectral lines with the known atomic spectra. As this can be done with many spectral lines, the redshift is by far the most precise quantity in cosmology. The cosmological redshift is typically taken as a measure of time; the larger $z$ the older was the source. It is of course not proportional to cosmological time, but as Eqs.~(\ref{eq:z}) show, directly related to the scale factor $a$ (and inversely proportional to $a$ for large $z$). 

\begin{figure}[h]
\begin{center}
\includegraphics[width=35pc]{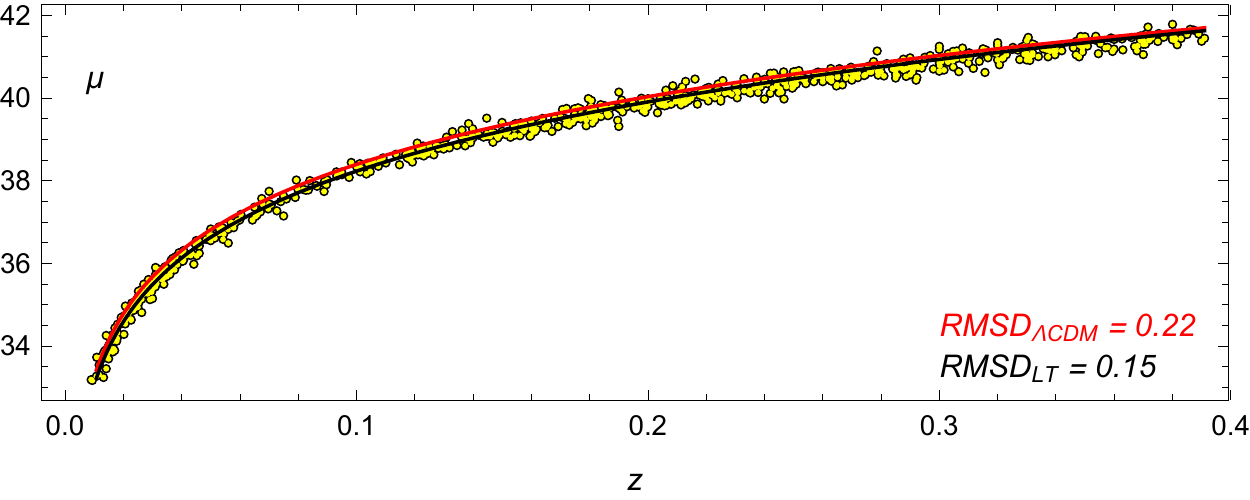}
\caption{
\small{
Hubble diagram. Logarithm $\mu$ of the luminosity distance versus redshift $z$ of supernova explosions from the Pantheon sample \cite{Scolnic}. The luminosity distance carries units and contains calibration factors that all appear as an offset in $\mu$ \cite{Scolnic}. The red curve shows the prediction from the $\Lambda$CDM model with parameters  [Eq.~(\ref{eq:parameters})] inferred from the correlations (Fig.~\ref{fig:correlations}) in the Cosmic Microwave Background. They originate from physics of the early universe ($z\sim10^3$) and one integration to the present (Sec.~4). The red curve fits the data well, but the black curve fits them even better. This curve was calculated using the Lifshitz theory of the quantum vacuum \cite{Berechya}. 
}
\label{fig:hubble}}
\end{center}
\end{figure}

Since the product $a\omega$ of scale factor and frequency is constant, we obtain by differentiation $\dot{\omega}/\omega=-\dot{a}/a$. Here and elsewhere a dot denotes differentiation with respect to cosmological time $t$. For light with $\mathrm{d}\ell=c\,\mathrm{d}t$ we thus get
%%%%%%
\begin{equation}
\mathrm{d}\omega = -\frac{H}{c}\,\mathrm{d}\ell 
\label{eq:hubblelaw}
\end{equation}
%%%%%%
with
%%%%%%
\begin{equation}
H = \frac{\dot{a}}{a} \,.
\label{eq:hubble}
\end{equation}
%%%%%%
Equation (\ref{eq:hubblelaw}) describes Hubble's law of increasing redshift with distance (decreasing frequency) and $H$ is called the Hubble parameter or Hubble constant in the early literature, although it is constant only for the case of exponential expansion. These days $H$ for $z=0$ (the present time) is called the Hubble constant $H_0$. By finding the relation (\ref{eq:hubblelaw}) between the redshift and the distance in astronomical data, Hubble discovered the expansion of the universe \cite{Hubble}. Hubble's law is still of vital importance today, in particular in the Hubble tension \cite{DiValentino}.  Equation (\ref{eq:hubblelaw}) is written in differential form, which is sufficient for the vicinity of our galaxy as the order of magnitude for the current $H$ is $10^{-10}$ per year. For more distant sources we need to integrate Hubble's law. The fundamental distance to some astronomical object corresponds to the time it takes for light to travel from there, the conformal time $\tau$. Multiplied by $c$ it is called comoving distance \cite{Dodelson}. For the luminosity distance $c\tau$ is multiplied by $1+z$ (divided by $a$) and for the angular distance it is divided by $1+z$ (multiplied by $a$) \cite{Dodelson}. For $\tau$ we have in terms of the Hubble parameter:
%%%%%%
\begin{equation}
\tau = \int \frac{\mathrm{d}a}{a^2 H} \,.
\label{eq:tauhubble}
\end{equation}
%%%%%%
In view of the vast cosmological distances, the logarithm of the luminosity distance is taken as a measure of distance and this logarithm as a function of redshift is drawn in a diagram called the Hubble diagram (for an example see Fig.~\ref{fig:hubble}). The Hubble diagram represents an integrated form of Hubble's law.

%%%
%%%

As Hubble had discovered \cite{Hubble} the universe is expanding, but the comoving coordinates $\bm{r}$ of, say, two objects remain the same. What expands is their (infinitesimal) distance $a\,\mathrm{d}r$. Let us introduce a new set of spatial coordinates $\bm{x}$ where we incorporate the expansion in space:
%%%%%%
\begin{equation}
\bm{x} = a \bm{r}\,.
\label{eq:x}
\end{equation}
%%%%%%
Their coordinate distance $x$ directly gives the proper distance $ar$. The coordinate origin we place at the position of the observer where $\bm{x}=\bm{0}=\bm{r}$. The $\bm{x}$--coordinates are fixed to the observer and are no longer comoving with the universe. In these fixed coordinates, space will not be homogeneous. We obtain from $\mathrm{d}\bm{x}=a\,\mathrm{d}\bm{r}+H\bm{x}\,\mathrm{d}t$ and Eq.~(\ref{eq:metric}) the transformed metric:
%%%%%%
\begin{equation}
\mathrm{d}s^2 = c^2\mathrm{d}t^2 - \mathrm{d}\bm{x}'^2 \,,\quad \mathrm{d}\bm{x}'=\mathrm{d}\bm{x}-\bm{u}\mathrm{d}t
\label{eq:metricflow}
\end{equation}
%%%%%%
with
%%%%%%
\begin{equation}
\bm{u}=H \bm{x} \,.
\label{eq:hubbleflow}
\end{equation}
%%%%%%
The $\bm{x}'$ we view as a new coordinate frame, but we can only do this locally and for a given moment in time. In this local frame the metric is exactly the same as in Minkowski space. The transformation from $\bm{x}$ to $\bm{x}'$ is a Galilei transformation with velocity $\bm{u}$; the $\bm{x}'$ are moving coordinates: space appears to flow. In the comoving $\bm{r}$--coordinates the positions of two objects are fixed, and in the fixed $\bm{x}$--coordinates the expanding universe pulls them apart (very slowly, with $H\sim 10^{-10}$ per year). The velocity $\bm{u}$ of Eq.~(\ref{eq:hubbleflow}) points in radial direction from the observer and grows linearly with distance $x$. The velocity profile is called the Hubble flow and the proportionality between velocity and distance is another Hubble law.

\begin{figure}[h]
\begin{center}
\includegraphics[width=15.0pc]{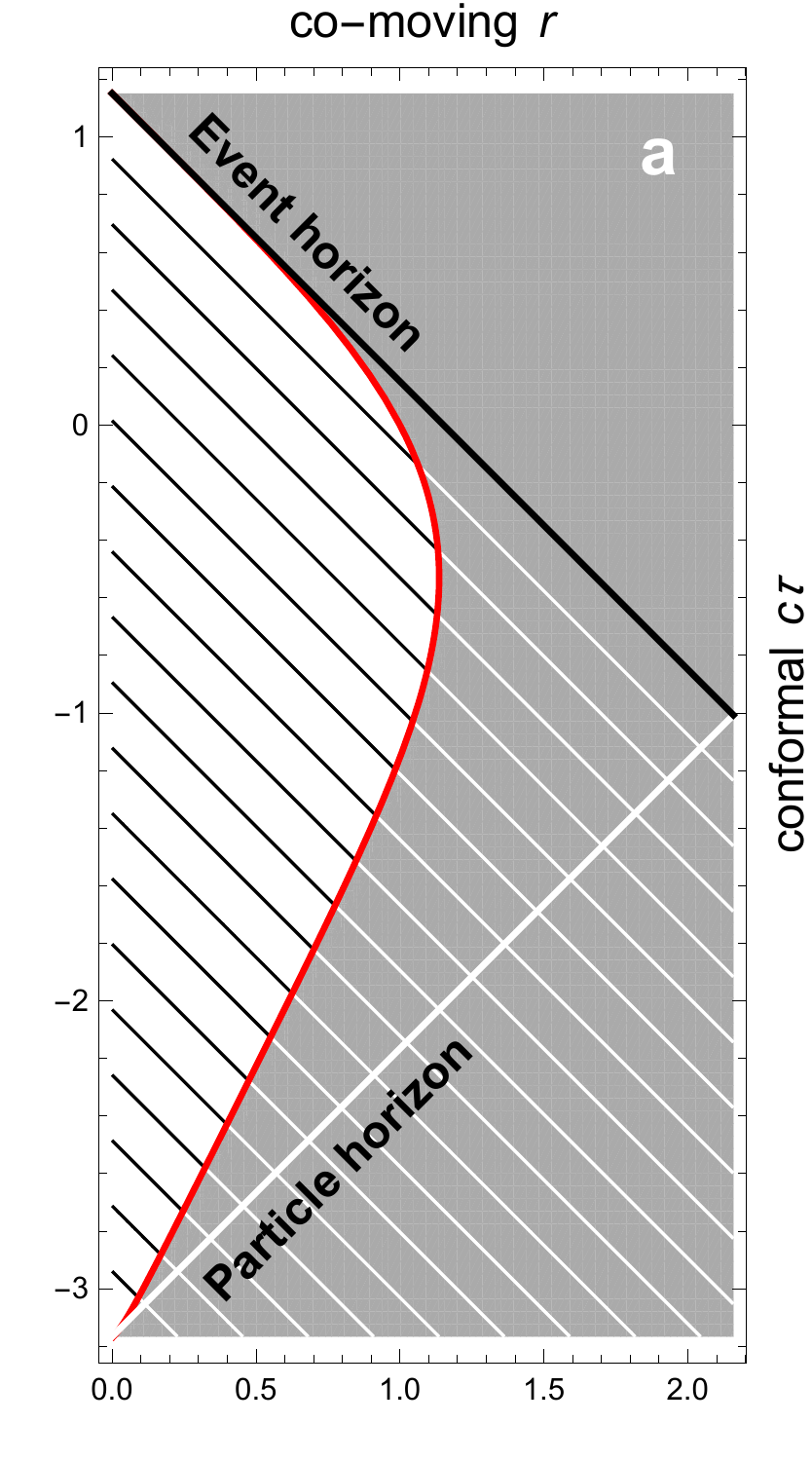}
\includegraphics[width=18.0pc]{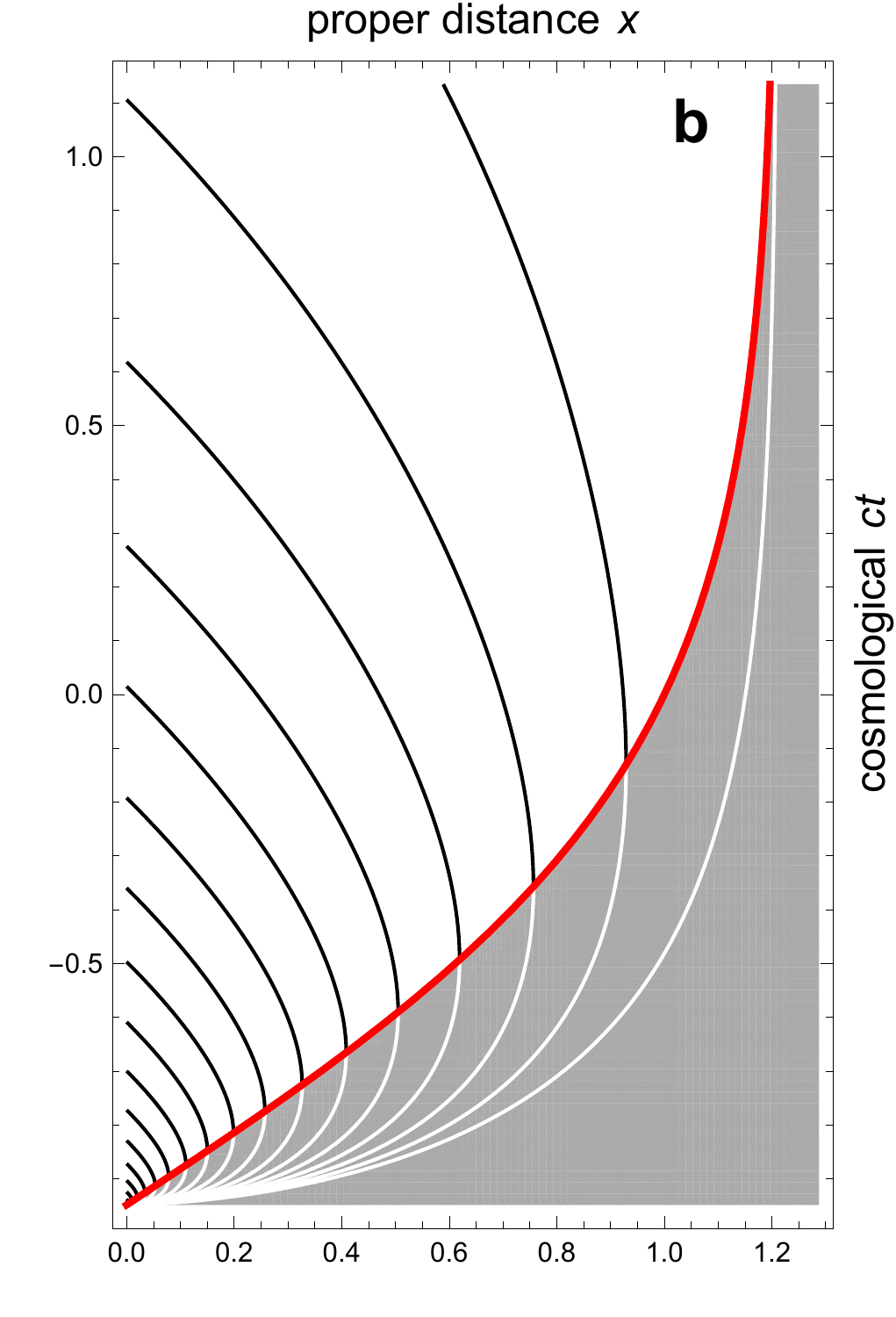}
\caption{
\small{
Horizons. Space--time diagrams of the universe with parameters (\ref{eq:parameters}) in units of $1/H_0$ for time and $c/H_0$ for space, at the present $t_0=\tau_0=0$. The red curves show the cosmological horizon, Eq.~(\ref{eq:horizon}) where for an observer at the origin the expansion velocity reaches $c$. The inside of the horizon is white, the outside grey. {\bf a}: in conformal time $\tau$ and comoving space $r$ light propagates along straight $45^\circ$ lines (white and black lines, white outside the horizon and black inside). {\bf b}: in cosmological time $t$ and proper distance $x$ the same light rays as in {\bf a} come to a brief standstill at the horizon (white and black curves) sufficient for generating Gibbons--Hawking radiation \cite{EPL}. 
}
\label{fig:horizons}}
\end{center}
\end{figure}

The Hubble flow increases slowly with distance, but as flat space is infinite, it will eventually exceed the speed of light. The place where the expansion velocity $\bm{u}$ reaches $c$ is called the cosmological horizon \cite{Harrison}. The horizon is the surface of a sphere of radius 
%%%%%%
\begin{equation}
x_H=\frac{c}{H} \,,\quad r_H = \frac{c}{aH}
\label{eq:horizon}
\end{equation}
%%%%%%
in fixed and comoving coordinates, respectively. Any point in the universe is surrounded by a cosmological horizon, because we can move the origin of the spatial coordinates to any point. Note that the cosmological horizon is not necessarily an event horizon \cite{Confusion}, because light may cross it in general (Fig.~\ref{fig:horizons}a). For example, the CMB has definitely crossed our horizon (see Sec.~3). The horizon does prevent light from entering if it divides the flow of light into light reaching the observer and light passing by, which is only possible if the horizon is light--like itself. In conformal time and comoving radius (Fig.~\ref{fig:horizons}a) light to the left of the event horizon reaches the observer, whereas on the right it never will (as the universe ``ends'' in conformal time before that could happen). The cosmological horizon is light--like for $\mathrm{d}r_H=-c\,\mathrm{d}\tau$ and $r_H=c/(a H)$. We see from Eq.~(\ref{eq:tauhubble}) that $H$ must be constant for this to be true. Only in a phase of exponential expansion (in de Sitter space \cite{deSitter}) the cosmological horizon is an event horizon. 

%%%
%%%

The event horizon of the black hole has entropy \cite{Bekenstein} and emits Hawking radiation \cite{Hawking,Brout}. Gibbons and Hawking \cite{GH} predicted that the cosmological event horizon of the exponentially expanding universe radiates as well, with temperature $T$ given by the formula
%%%%%%
\begin{equation}
k_\mathrm{B} T = \frac{\hbar H}{2\pi} 
\label{eq:gh}
\end{equation}
%%%%%%
where $\hbar$ denotes the reduced Planck constant and $k_\mathrm{B}$ Boltzmann's constant. Analogues of the event horizon have been discussed and made in the laboratory \cite{Volovik,Visser,Unsch,Faccionotes}. In fluids \cite{Visser}, Hawking radiation is emitted whenever the flow speed $u$ reaches the speed of the waves $c$ (in the regime of weak dispersion). Supported by the experimental and theoretical evidence for analogues of the event horizon, one can therefore generalize Gibbon's and Hawking's result for exponential expansion to an arbitrary expansion $a(t)$ \cite{Annals}. An exact calculation without the use of analogues corroborates this argument \cite{EPL}. Furthermore, Gibbons--Hawking radiation \cite{GH} and its generalization \cite{Annals,EPL} is related \cite{EPL,Good} to the dynamical Casimir effect \cite{SchwingerDCE,Mendonca1,Mendonca2,Dodonov} that has been experimentally observed \cite{Wilson,Hakonen,Veccoli}. In the dynamical Casimir effect one typically pays attention to the fact that photons are produced from the quantum vacuum. But in addition to particle creation, the dynamical Casimir effect also modifies the quantum noise of the vacuum. Without it, the vacuum noise would be exactly the same as in Minkowski space where there are no vacuum forces. Due to the Gibbons--Hawking effect, the renormalized vacuum noise becomes cosmological relevant in principle. It also seems relevant in practice \cite{Annals,London} although the Gibbons--Hawking temperature (\ref{eq:gh}) is astronomically small ($2\times10^{-29}\mathrm{K}$ at the present time).

\section{Cosmic dynamics}

So far we considered only the kinematic aspects of cosmology, let us now add the dynamics. Two forces determine the cosmic evolution: gravity \cite{LL2} and fluid mechanics \cite{LL6}. Averaged over cosmological scales,  all the intricate structures of the physical world blend into a homogeneous fluid. Galaxies reduce to specks of ``dust'' and the universe is uniformly filled with radiation of photons and neutrinos. Each cosmic fluid has its energy density $\varepsilon$ and pressure $p$. There are several such fluids relevant to the cosmic evolution, let us first assume only one fluid for simplicity. On average, this fluid is comoving with the universe, for otherwise any macroscopic velocity $\bm{v}$ would distinguish a direction and hence violate the cosmological principle of homogeneity and isotropy. We postulate that the cosmic fluid expands adiabatically, {\it i.e.}\ entropy is conserved. In this case we have from thermodynamics \cite{LL5} for the energy $E$ and volume $V$ the relation $\mathrm{d}E=-p\,\mathrm{d}V$. Writing $E=\varepsilon V$ and using the fact that the volume goes with $a^3$ we get \cite{LL2}:
%%%%%%
\begin{equation}
\mathrm{d}\varepsilon = -3(\varepsilon+p)\frac{\mathrm{d} a}{a} \quad\mbox{or}\quad
\dot{\varepsilon}=-3H(\varepsilon+p) \,.
\label{eq:f2}
\end{equation}
%%%%%%
Now for gravity. The energy density is related to the mass density $\rho$ by Einstein's 
%%%%%%
\begin{equation}
\varepsilon=\rho c^2 \,.
\label{eq:mc2}
\end{equation}
%%%%%%
This mass density generates a gravitational field such that the expansion obeys the Friedman equation \cite{LL2}
%%%%%%
\begin{equation}
H^2 = \frac{8\pi G}{3}\rho
\label{eq:f1}
\end{equation}
%%%%%%
in the (realistic) case of zero spatial curvature. The quantity $G$ denotes Newton's gravitational constant. The Friedman equation (\ref{eq:f1}) and the thermodynamic relation (\ref{eq:f2}) --- also known as the first and the second Friedman equations --- solve Einstein's field equations for the metric of Eq.~(\ref{eq:metric}) \cite{LL2}. 

\begin{figure}[h]
\begin{center}
\includegraphics[width=16.0pc]{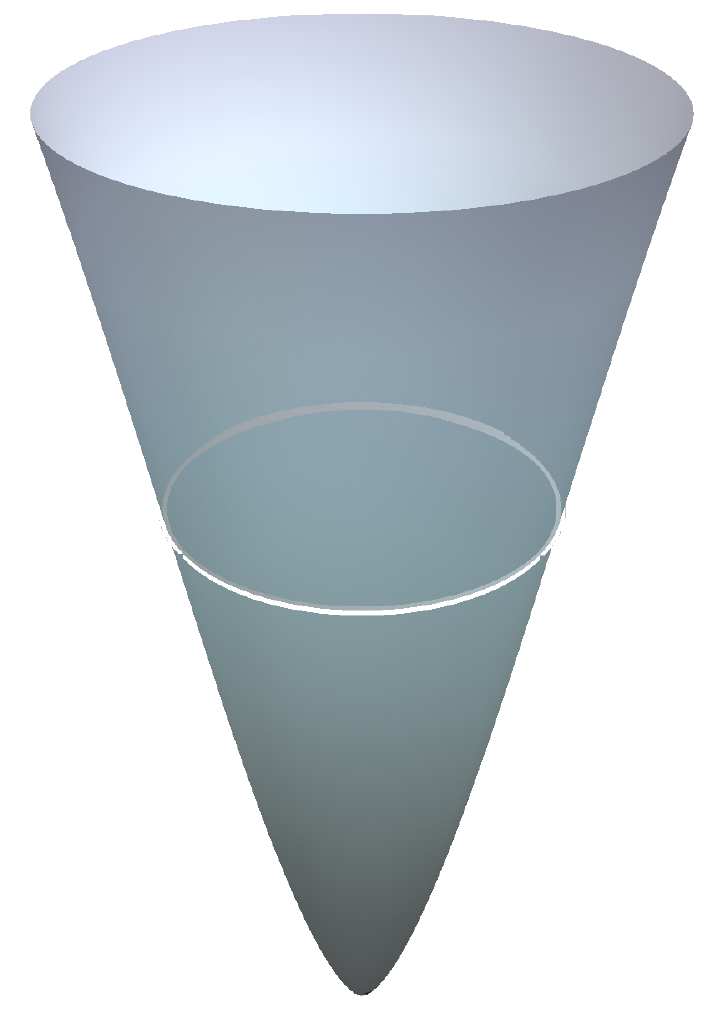}
\caption{
\small{
Cosmic expansion. Space--time diagram of the universe with parameters (\ref{eq:parameters}). Cosmological time $t$ runs in vertical direction from the big bang to the present time. In the horizontal direction the radius of the figure represents the scale factor $a(t)$. Initially, the bell--shaped figure bends inwards as gravity de--accelerates the expansion. However, eventually (white ring) the bell begins to curve outwards:  $\Lambda$ takes over and accelerates the expansion of the universe.}
\label{fig:expansion}}
\end{center}
\end{figure}

The cosmic dynamics is a consequence of Einstein's general relativity, but it (almost) follows from Newtonian gravity \cite{Milne}. To see this, differentiate the Friedman equation (\ref{eq:f1}) with respect to cosmological time, use the thermodynamic relation (\ref{eq:f2}) with the energy--mass relationship (\ref{eq:mc2}) and the mathematical identity $\dot{H}=\ddot{a}/a-H^2$ with Eq.~(\ref{eq:f1}) for $H^2$ to obtain
%%%%%%
\begin{equation}
\frac{\ddot{a}}{a} = - \frac{4\pi G}{3} \left(\rho + 3\frac{p}{c^2}\right) .
\label{eq:newton}
\end{equation}
%%%%%%
This dynamical equation of the expanding universe is sometimes also called Friedman equation and also the acceleration equation. In condensed matter, analogues of Eq.~(\ref{eq:newton}) appear in Coulomb explosions where electrostatics plays the role of gravity \cite{Kolo1,Kolo2,Kolo3}. In cosmology, Eq.~(\ref{eq:newton}) has a Newtonian interpretation if we think of the scale factor not as a measure of spatial distance but rather as the distance between points in absolute space at the same absolute time. Relativity only adds two features, one is obvious from Eq.~(\ref{eq:newton}) and the other subtle. The subtle feature is expressed in Eq.~(\ref{eq:mc2}): not only rest mass gravitates, but all energy contributes to the mass according to $E=mc^2$. Equation (\ref{eq:newton}) then states that pressure also contributes to gravity. As pressure is exerted from all 3 sides the contribution goes with $3p c^{-2}$ such that the effective gravitational mass density is $\rho_\mathrm{eff}=\rho+3p c^{-2}$. We may read Eq.~(\ref{eq:newton}) as the equation of a harmonic oscillator with spring constant given by the volume of the unit sphere $4\pi/3$ times $G\rho_\mathrm{eff}$. Now, the Newtonian gravitational force inside a homogeneous fluid is exactly the force of this harmonic oscillator (which is a simple consequence of Gauss' law). The universe thus expands as it should in Newtonian dynamics. However, if the pressure $p$ is negative and in the order of $\rho c^2$ the spring constant changes sign. In this case, gravity is no longer an attractive force, but becomes repulsive; instead of slowing down the expansion of the universe, gravity accelerates it. As a matter of fact  \cite{CMBPlanck,Super1,Super2}, this is the case at the present stage of expansion (Fig.~\ref{fig:expansion}) and it makes the case for the cosmological constant, as we will see next.

%%%
%%%

But before discussing the actual cosmic evolution, note that we can take the Newton equation (\ref{eq:newton}) as the starting point and integrate it, using the thermodynamic relation (\ref{eq:f2}). We obtain the Friedman equation (\ref{eq:f1}) as an integral of the Newton equation, apart from an additional term coming from the integration constant $K$. We put it on the left--hand side and write:
 %%%%%%
\begin{equation}
\frac{\dot{a}^2}{a^2} + \frac{K}{a^2} = \frac{8\pi G}{3}\rho \quad\mbox{with}\quad 
K=\frac{\mathrm{k} }{t_K^2}
\label{eq:fk}
\end{equation}
%%%%%%
in terms of a time constant $t_K$ and $\mathrm{k}\in\{-1,0,+1\}$. The integration constant $K$ depends on the initial velocity at $a=0$. In the Einsteinian picture the Newtonian initial velocity finds a geometric interpretation: it turns out \cite{LL2} to describe spatial curvature. Space with $\mathrm{k}=+1$ corresponds to the 3D surface of the 4D hypersphere, whereas space with $\mathrm{k}=-1$ we may picture as the surface of the pseudo-hypersphere \cite{LL2}. The radius of curvature is given by $ct_K$. For $\mathrm{k}=0$ space is flat (but space--time is curved due to the cosmic evolution in time \cite{LL2}). Flat space and space with constant negative curvature are open spaces infinitely extended, whereas space with positive constant curvature, being the surface of the hypersphere, is closed. Space needs to be curved in one way or the other depending on the initial real number $K$ --- the chances of exactly zero curvature are exactly zero. But it is not known yet which way the curvature went and so whether we inhabit an open or a closed universe. Most probably, just after the big bang cosmic inflation \cite{Inflation} has exponentially enlarged the radius of curvature $ct_K$ such that the curvature term $Ka^{-2}$ has become too small to be measurable at the moment. Let us put $K=0$ and turn to the right--hand side of the Friedman equation (\ref{eq:f1}).

According to the cosmological standard model \cite{CMBPlanck} there are three contributions to the mass density: radiation R (photons and neutrinos), matter M (baryonic and dark) and the cosmological constant $\Lambda$. Let us deduce how their mass densities depend on $a$. First, any density must go like $a^{-3}$ in the expanding universe. The energy $\hbar\omega$ of radiation of frequency $\omega$ falls with $a^{-1}$ as $a\omega=\mathrm{const}$. So the total radiation energy density $\varepsilon_\mathrm{R}$ goes like $a^{-4}$ and, according to Eq.~(\ref{eq:mc2}), so does the mass density $\rho_\mathrm{R}$. For $\varepsilon_\mathrm{R}\propto a^{-4}$ we obtain from the thermodynamic relation (\ref{eq:f2}):
%%%%%%
\begin{equation}
p_\mathrm{R} = \frac{\varepsilon_\mathrm{R}}{3} \,.
\label{eq:radiationpressure}
\end{equation}
%%%%%%
For non--relativistic matter the pressure is completely negligible in comparison with the rest--energy density. We thus put
%%%%%%
\begin{equation}
p_\mathrm{M} = 0 \,,
\label{eq:matterpressure}
\end{equation}
%%%%%%
and obtain from Eq.~(\ref{eq:f2}) that $\rho_\mathrm{M}\propto a^{-3}$. This is consistent with the notion that the mass of non--relativistic matter equals the rest mass and so does not change during cosmic expansion, which implies that the mass density goes with $a^{-3}$. For $\Lambda$ we postulate that $\varepsilon_\Lambda=\mathrm{const}>0$. From the thermodynamic relation (\ref{eq:f2}) follows that this is only possible if
%%%%%%
\begin{equation}
p_\Lambda= -\varepsilon_\Lambda \,.
\label{eq:plambda}
\end{equation}
%%%%%%
From the Newton equation (\ref{eq:newton}) and Einstein's relation (\ref{eq:mc2}) then follows that $\Lambda$ causes a repulsive contribution to gravity. Without $\Lambda$ the spatial curvature $K$ would ultimately dictate the fate of the universe, because its contribution goes like $a^{-2}$ and all other contributions to the total mass density are diluted faster. The right--hand side of the Friedman equation (\ref{eq:fk}) would go faster to zero than $\mathrm{k}/(t_K a)^2$ and so $\dot{a}^2+\mathrm{k}/t_K^2\sim\mathrm{const}$, which implies that $a\propto\sin[(t-t_0)/t_K]$ for $\mathrm{k}=+1$ and $a\propto\sinh[(t-t_0)/t_K]$ for $\mathrm{k}=-1$. Consequently, for positive curvature, the universe reaches a maximal size at a cosmological time in the order of $t_K$ after the big bang, and then shrinks until the ``big crunch''. A universe of positive spatial curvature would therefore be closed not only in space but also in time, whereas for the open universe of negative spatial curvature the expansion would go on forever. In the Newtonian interpretation of the cosmic dynamics this is a feature familiar from gravitational escape. Here the curvature corresponds to the initial velocity. For $\mathrm{k}>0$ the left--hand side of the Friedman equation (\ref{eq:fk}) is reduced below the escape velocity and for $\mathrm{k}<0$ it is enhanced. The density for $\mathrm{k}=0$ is called the critical density and it corresponds to the exact escape velocity. However, with $\varepsilon_\Lambda>0$ the cosmological constant ultimately becomes dominant, causing the expansion to accelerate regardless of spatial curvature. Whatever happens, gravitational repulsion eventually takes over (Fig.~\ref{fig:expansion}).

The details depend on the numbers. The proportions of the three constituents of the cosmological standard model, the $\Lambda$CDM model, are typically expressed by writing the Friedman equation (\ref{eq:f1}) as
%%%%%%
\begin{equation}
H^2 = H_0^2 \left(\frac{\Omega_\mathrm{R}}{a^4}+\frac{\Omega_\mathrm{M}}{a^3} +\Omega_\Lambda\right) 
\label{eq:lambdacdm}
\end{equation}
%%%%%%
where the $\Omega_m$ denote the relative weights of the various mass--density contributions at the present time. As $a(t_0)=1$ we thus have
%%%%%%
\begin{equation}
\sum_m \Omega_m = 1 \,.
\label{eq:omegasum}
\end{equation}
%%%%%%
The prefactor $H_0$ gives the Hubble parameter at the present time, the Hubble constant, provided of course that Eq.~(\ref{eq:lambdacdm}) holds exactly, which is almost exactly true, but not quite \cite{DiValentino}. We will discuss the main weakness of the $\Lambda$CDM model in Sec.~4. In any case, the model--independent quantities are the mass densities in the Friedman equation (\ref{eq:f1}) of form (\ref{eq:lambdacdm}):
%%%%%%
\begin{equation}
H_0^2\Omega_m = \left.\frac{8\pi G}{3}\rho_m\right|_{a=1} \,.
\label{eq:rhos}
\end{equation}
%%%%%%
The contribution of radiation is given by the Stefan--Boltzmann law \cite{Weinberg} for the CMB with average temperature $T_0$ at the present time:
%%%%%%
\begin{equation}
\frac{3H_0^2\, \Omega_\mathrm{R} }{8\pi G} = \rho_\mathrm{R} = \eta_\mathrm{R}\,\frac{\pi^2\,(k_\mathrm{B}T_0)^4}{15\,c^5\,\hbar^3} \,,\quad \eta_\mathrm{R} = 1+3\, \frac{7}{8} \left(\frac{4}{11}\right)^{4/3} 
\label{eq:stefanboltzmann}
\end{equation}
%%%%%%
taking into account the fermionic factor $7/8$ for the three neutrino flavours and the feature that cosmic neutrinos are $(4/11)^{1/3}$ colder than the photonic CMB \cite{Weinberg}. The average temperature $T_0$ of the present CMB is measured with high precision. The other $\Omega_m$ including a potential curvature contribution and $H_0$ are obtained from the fluctuations of the CMB (Sec.~4). The numbers are \cite{CMBPlanck}:
%%%%%%
\begin{eqnarray}
H_0 &=& (67.36 \pm 0.54) \,\frac{\mathrm{km}/\mathrm{s}}{\mathrm{Mpc}} \,,\nonumber\\
\Omega_\mathrm{R} &=& 0.925\times10^{-4} \,,\nonumber\\
\Omega_\mathrm{M} &=& 0.3153 \pm 0.0073 \,,\nonumber\\
\Omega_\Lambda &=& 0.6847 \pm 0.0073 \,.
\label{eq:parameters}
\end{eqnarray}
%%%%%%
We may also describe the spatial curvature as an $\Omega$--contribution to the right--hand side of Eq.~(\ref{eq:lambdacdm}). From Eq.~(\ref{eq:fk}) follows the curvature contribution $\Omega_K a^{-2}$, but $\Omega_K$ is approximately zero with currently best \cite{CMBPlanck} bound of $|\Omega_K|<2\times 10^{-3}$. 

The numbers (\ref{eq:parameters}) tell that $\Lambda$ makes up roughly $70\%$ of the current mass density. The units of the Hubble constant $H_0$ indicate that the Hubble flow of Eq.~(\ref{eq:hubbleflow}) only becomes macroscopically fast on the $\mathrm{Mpc}$ ($\mathrm{Mly}$) scale. The current cosmological horizon $c/H_0$ surrounds us $14.6 \mathrm{Gly}$ away. With the model (\ref{eq:lambdacdm}) and the parameters (\ref{eq:parameters}) we can easily calculate the age of the universe: the cosmological time $t_0$ of the big bang
%%%%%%
\begin{equation}
t_0 = \int^1_0 \frac{\mathrm{d}a}{aH} \,.
\label{eq:age}
\end{equation}
%%%%%%
From the parameters (\ref{eq:parameters}) we obtain the value $t_0=13.8\mathrm{Gy}$. For the conformal time of Eq.~(\ref{eq:tauhubble}) we get $\tau_0=46.2\mathrm{Gy}$. Now, $c\tau$ describes the distance light has travelled. This distance in light--years is more than a factor of three longer than the age of the universe in years. Since the CMB was released shortly after the big bang (shortly on the $\mathrm{Gy}$ scale) this is the distance from which the CMB reaches us. It is significantly larger than the present cosmological horizon. We should however compare this distance with the past horizon we are experiencing, not with the present one we are establishing. For determining the past horizon $r_0$ we follow the radiation back as $r_0=-c\tau$ in terms of the conformal time $\tau$ and calculate the moment in $\tau$ when $c\tau$ intersects the horizon in the space--time diagram (Fig.~\ref{fig:horizons}a). For this we tabulate the values of $\tau(a)$ and solve for $\tau=-1/(aH)$. The result $r_0=15.2\mathrm{Gly}$ confirms that the CMB has reached us from beyond the horizon. Furthermore, the redshift of the horizon is $z=1/a-1=1.59$, and one can observe galaxies with larger redshifts, which proves their light has crossed the horizon, too. Yet beyond the horizon the universe expands faster than light can travel. So why is light not stopped there? Figure~\ref{fig:horizons}b illustrates that at the horizon light stands still for the briefest of moments (in the fixed $\bm{x}$--coordinates) but then the horizon  moves outward and the light gets in. 

We may also use the parameters (\ref{eq:parameters}) of the $\Lambda$CDM model (\ref{eq:lambdacdm}) to estimate the characteristic scales for the succession of characteristic periods in the cosmic expansion. Initially, the cosmic evolution is dominated by radiation, as $\Omega_\mathrm{R}a^{-4}$ is the dominant term in Eq.~(\ref{eq:lambdacdm}) for $a\sim 0$, regardless how small $\Omega_\mathrm{R}$ is. At radiation--matter equality $\rho_\mathrm{R}=\rho_\mathrm{M}$, which is the case for 
%%%%%%
\begin{equation}
a_\mathrm{eq} = \frac{\Omega_\mathrm{R}}{\Omega_\mathrm{M}} \,,\quad
z_\mathrm{eq} = a_\mathrm{eq}^{-1}-1\,.
\label{eq:rm}
\end{equation}
%%%%%%
From the parameters (\ref{eq:parameters}) we get $z_\mathrm{eq}\approx 3\times 10^3$. The CMB is released at a redshift $z_\mathrm{*}\approx 10^3$ (Sec.~4). As the CMB is formed during an era when both matter and radiation are equally relevant, the fluctuations of the CMB contain sufficient information to retrieve the cosmic parameters (see Sec.~4 and Appendix A). At about the release of the CMB, matter dominates the cosmic evolution until the vacuum energy $\varepsilon_\Lambda$ becomes important and accelerates the expansion (Fig.~\ref{fig:expansion}). The acceleration changes sign when $\varepsilon+3p/c^2=0$ [Eq.~(\ref{eq:newton})]. As the radiation contributions are negligible, $\varepsilon_\mathrm{M}\propto \Omega_\mathrm{M}/a^3$ and $p_\mathrm{M}=0$, whereas $\varepsilon_\Lambda=-p_\Lambda\propto\Omega_\Lambda$, the accelerating phase began at
%%%%%%
\begin{equation}
a_\Lambda = \left(\frac{\Omega_\mathrm{M}}{2\Omega_\Lambda}\right)^{1/3} \,,\quad
z_\Lambda = a_\Lambda^{-1}-1\,.
\label{eq:lambdapoint}
\end{equation}
%%%%%%
From the cosmic parameters (\ref{eq:parameters}) we obtain the figure $z_\Lambda\approx 0.6$ that corresponds to the cosmological time $t_\Lambda=6.1\mathrm{Gy}$. The figure for $t_\Lambda$ indicates that the acceleration began well within the matter--$\Lambda$ era at about the time the solar system was formed. The moderate redshift $z_\Lambda$ shows that we are currently in the transition period from matter to vacuum domination. Solving in the Friedman equation (\ref{eq:f1}) for $\rho$ and putting $H=H_0$ we get the current average density:
%%%%%%
\begin{equation}
\rho_0 = \frac{3 H_0^2}{8\pi G} \approx 10^{-27} \mathrm{g}\,\mathrm{cm}^{-3} \,.
\label{eq:total}
\end{equation}
%%%%%%
This figure brings home how empty the universe is --- and $70\%$ of that figure appears to be made of vacuum energy. The figure vividly illustrates that the physics of the quantum vacuum matters in cosmology. 

\section{Cosmic Microwave Background}

\begin{figure}[h]
\begin{center}
\includegraphics[width=32.0pc]{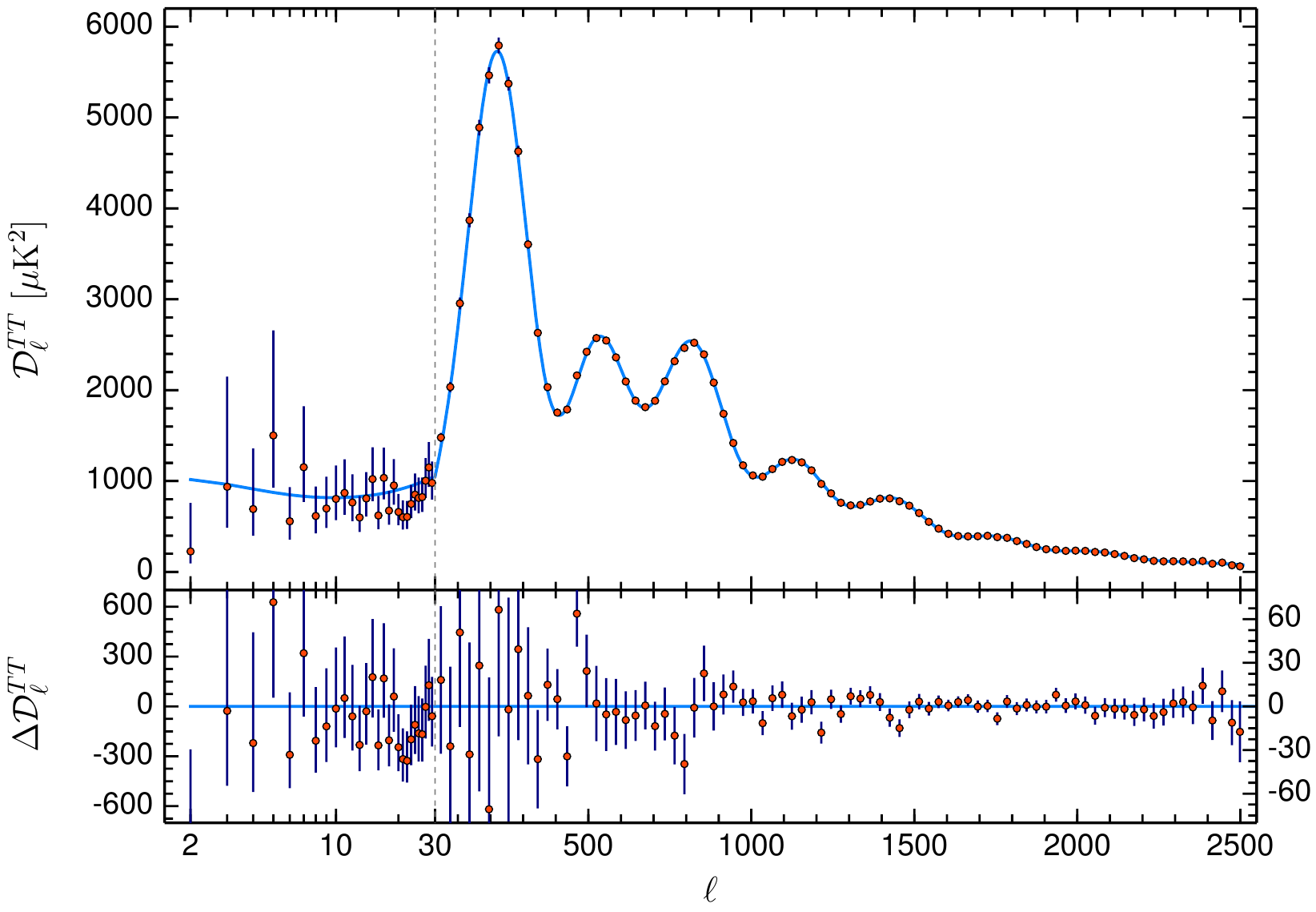}
\caption{
\small{
Correlations in the Cosmic Microwave Background (CMB). The figure, reproduced from Ref.~\cite{CMBPlanck}, shows the correlations (dots, with $1\sigma$ error bars) in the temperature fluctuations for the spherical harmonics with index $l$ (logarithmic scale for $l\le30$, then linear). The correlations are defined as ${\cal D}^{TT}_l = l(l+1)\,C_l$ with $C_l=(4\pi)^{-1}\int \delta T(\bm{n})\,\delta T(\bm{n}')\,P_l(\bm{n}\cdot\bm{n}')\, \mathrm{d}^2\bm{n}\,\mathrm{d}^2\bm{n}'$ for the angular directions $\bm{n}$ and in terms of the Legendre polynomials $P_l$. The solid curve is drawn from the prediction of the $\Lambda$CDM model with parameters (\ref{eq:parameters}); the lower panel shows the deviations from the curve. The nearly perfect fit strongly supports the $\Lambda$CDM model and allows for the precise determination of the cosmic parameters (\ref{eq:parameters}). 
}
\label{fig:correlations}}
\end{center}
\end{figure}

The Cosmic Microwave Background (CMB) is the electromagnetic part of the radiation filling the universe according to the Stefan--Boltzmann law (\ref{eq:stefanboltzmann}) with $\eta_\mathrm{R}=1$. It amounts to less than $10^{-4}$ of the total density (\ref{eq:total}) but it contains sufficient information to retrieve the cosmic parameters (\ref{eq:parameters}). This information is contained in the temperature $T_0$ and the temperature fluctuations $\delta T_0$ of the CMB (and also in its polarization \cite{CMBPlanck}). The CMB shows a nearly perfect thermal Planck spectrum, with hardly visible error bars, of the temperature \cite{Fixsen}: 
%%%%%%
\begin{equation}
T_0  = 2.7255\mathrm{K}\, .
\label{eq:temperature0}
\end{equation}
%%%%%%
During the cosmic evolution, the temperature of thermal electromagnetic radiation goes with the scale factor $a$ like the frequency $\omega$, because the Boltzmann factor of a thermal state depends on the ratio $\hbar\omega/(k_\mathrm{B}T)$ and the statistical state remains the same. Consequently, 
%%%%%%
\begin{equation}
T  = \frac{T_0}{a} \,.
\label{eq:temperature}
\end{equation}
%%%%%%
The CMB was released when the universe became transparent. Before that, the temperatures were so high that light and baryonic matter were strongly coupled in a plasma, and light was scattered on free electrons and highly excited atoms. Due to cosmic expansion $T$ fell according to Eq.~(\ref{eq:temperature}) until, finally, the plasma recombined at the temperature $T_\mathrm{*}$ \cite{Weinberg}. At the time $t_\mathrm{*}$ of last scattering, light was no longer interacting with baryonic matter, and was released. Relating $T_\mathrm{*}$ to the present, measured $T_0$ by Eq.~(\ref{eq:temperature}) gives the scale factor $a_\mathrm{*}$ and the redshift $z_\mathrm{*}$ of last scattering. The refined, fitted values are \cite{CMBPlanck}:
%%%%%%
\begin{equation}
a_\mathrm{*} = 0.9166 \times 10^{-3} \,,\quad z_\mathrm{*} = 1090 \,.
\label{eq:zL}
\end{equation}
%%%%%%
After the time of last scattering the electromagnetic radiation freely propagated while being redshifted to the microwave part of the spectrum, filling the universe with a uniform microwave background of temperature $T_0$. Yet the temperature varies in different directions by about $10^{-5}T_0$ (with $18\mathrm{\mu K}$ variance \cite{Wright}). This variation is random, fluctuating  --- in space, not in time (on Human time scales). The fluctuations appear as variations of $T_0$ in different directions of the sky (after having subtracted the Doppler effect of our motion relative to the average CMB). But although each fluctuation is random, the fluctuations are statistically correlated. Their averages vanish, but not their variances, and the variances form an undulating curve (Fig.~\ref{fig:correlations}).

%%%
%%%

\begin{figure}[h]
\begin{center}
\includegraphics[width=20.0pc]{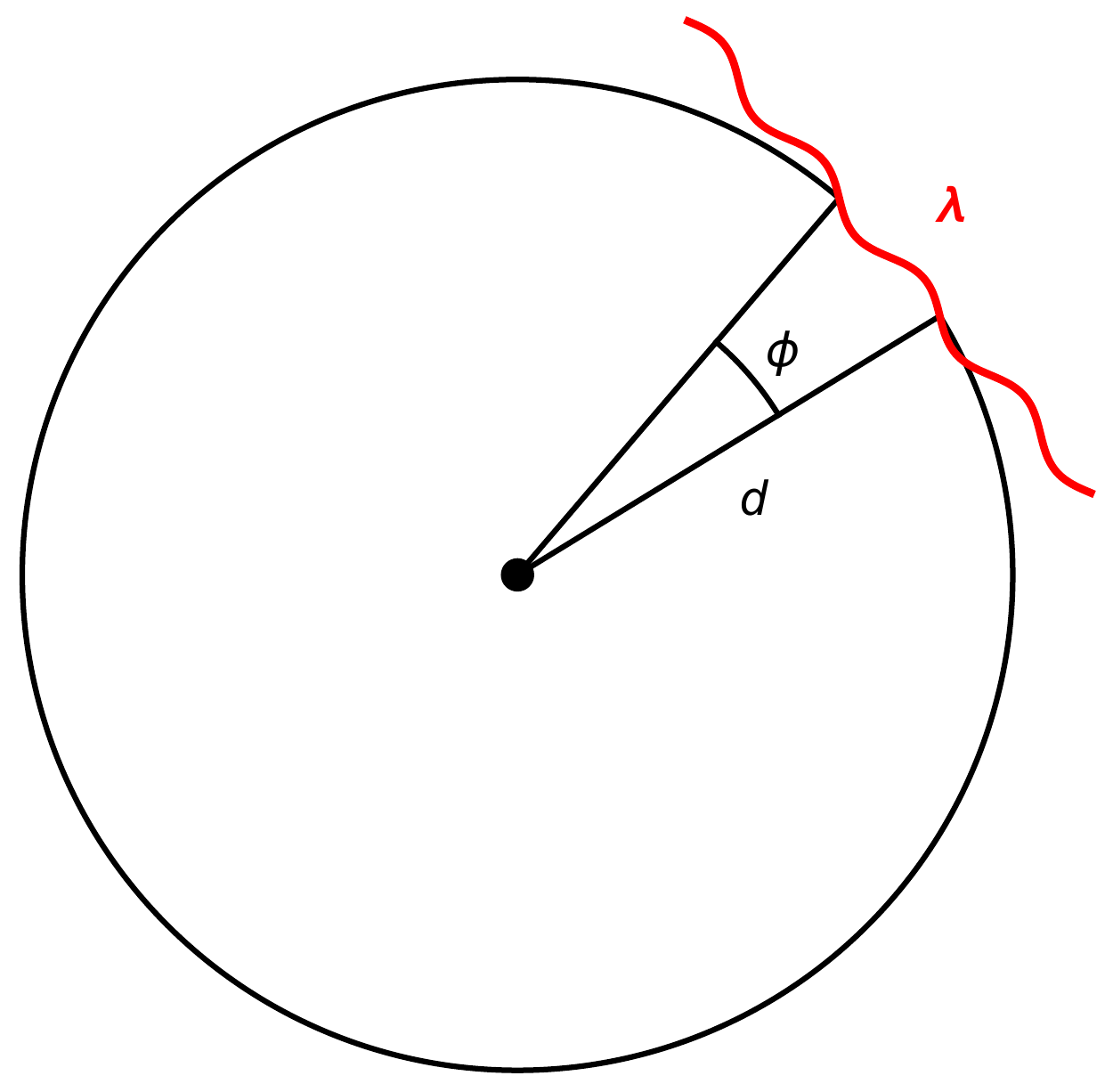}
\caption{
\small{
Angle, wavelength and distance. The CMB reaches us (central dot) from all directions. The radiation we receive today was released at the time $t_\mathrm{*}$ of last scattering. Since then it has traveled along straight lines in conformal time $\tau$ and comoving space (Fig.~\ref{fig:horizons}) the distance $d=c\tau_\mathrm{*}$. Acoustic waves (wavy line) modulate the radiation, causing characteristic correlations in the fluctuations (Fig.~\ref{fig:correlations}). The wavelength $\lambda$ of such an acoustic wave appears under the angle $\phi=\lambda/d$ for large $d$. Here $\lambda$ refers to the difference in comoving coordinates where $\mathrm{d}\ell=a\,\mathrm{d}r$; the actual wavelength is $a_\mathrm{*}\lambda$.}
\label{fig:angle}}
\end{center}
\end{figure}

This is a typical wave phenomenon. Wave noise is organised, because a random wave is not random in its space--time structure, but only in its overall amplitude and phase. Suppose plane waves of wave vectors $\bm{k}$ are randomly seeded with a well--defined initial noise depending on $k=|\bm{k}|$ (because of isotropy). Each $k$--component propagates with a different wave velocity and arrives with a different phase at some final time, here $t_\mathrm{*}$. Depending on this phase, the correlations get undulated. The CMB probes these undulating correlations. They appear as correlations in the spherical harmonics of the radiation pattern \cite{Weinberg}. To put it simply, the wave correlations appear as angle correlations by relating the angle $\phi$ to the wavelength $\lambda$ and distance $d$ (Fig.~\ref{fig:angle}) for small $\phi$:
%%%%%%
\begin{equation}
\phi = \frac{\lambda}{d} \,.
\label{eq:angle}
\end{equation}
%%%%%% 

%%%
%%%

\begin{figure}[h]
\begin{center}
\includegraphics[width=25.0pc]{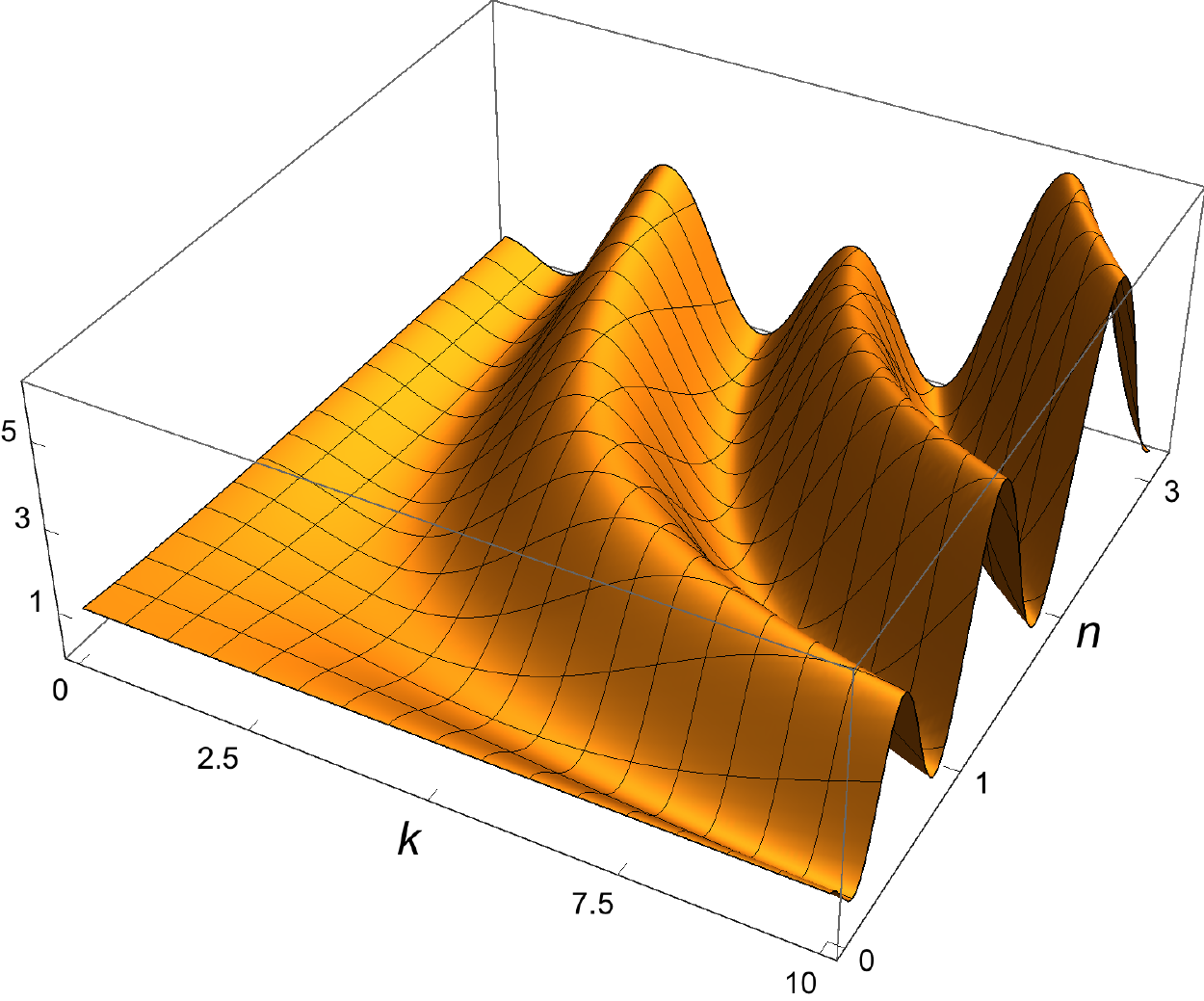}
\caption{
\small{
Cosmic sound. Temperature variance $(\delta T/T)^2$ in arbitrary units caused by acoustic waves with wavenumber $k$ propagating in the expanding universe with scale factor $n$ (all in scaled units). As the waves are seeded with equal amplitude at $n=0$ the initial $\delta T/T$ is uniform. For $n>0$ the waves oscillate with phase proportional to $k$ and the integrated speed of sound, Eq.~(\ref{eq:phase}). The picture shows how the temperature variance gets modulated due to the varying phase. After the release of the CMB at $n_\mathrm{*}=3.17$ the imprinted  $(\delta T/T)^2$ remains the same as $T$ falls with scale factor. The wavenumber $k$ appears (Fig.~\ref{fig:angle}) as an angular momentum $l$ in the correlation pattern (Fig.~\ref{fig:correlations}). The figure shows the first three peaks in $(\delta T/T)^2$ as function of $k$. In the actual correlations (Fig.~\ref{fig:correlations}) their amplitude gets damped with increasing $k$ due to diffusion damping (Silk damping \cite{Weinberg, Silk}). The temperate deviation was calculated [Eq.~(\ref{eq:deltaT})] after solving numerically the full fluid--mechanical equations of motions (Secs.~A.4 and A.5).
}
\label{fig:waves}}
\end{center}
\end{figure}

The fluctuations of the CMB bear all the markings of wave noise, but of which waves? The CMB reaches us from $40\mathrm{Gly}$ away (Sec.~3). For structures with angles of, say, around $4\times 10^{-3}\pi$ their wavelenghts must be in the order of $10\mathrm{Kly}$ taking the smaller scale  $a_\mathrm{*}$ at the time of last scattering into account. These astronomically large waves are acoustic waves: sound \cite{LL6}. A sound wave is made by two things: pressure and inertia. For the CMB, light provided the pressure and baryonic matter (plus light) the inertia (Appendix A). These sound waves of light and matter propagate on the background of the expanding universe. The waves are also modulated by the gravity of their own density variations and by the density variations of the other ingredients of the cosmic mixture, neutrinos and dark matter (Appendix A). The initial random seed of unstructured noise is believed to be vacuum noise parametrically amplified during inflation \cite{Inflation} just before the $\Lambda$CDM evolution of Eq.~(\ref{eq:lambdacdm}) started. The final, released waves imprint temperature fluctuations in the CMB due to their intrinsic temperature variations and due to their local gravitational redshifts (Fig.~\ref{fig:waves}, Appendix A). 

%%%
%%%

Despite all these complexities, we will only need some general scaling arguments to understand how the cosmic parameters (\ref{eq:parameters}) are extracted in principle. In practice \cite{CMBPlanck,Dodelson}, a more sophisticated fitting procedure is used for better accuracy. The problem simplifies thanks to a separation of scales. We see from Eq.~(\ref{eq:zL}) that the release of the CMB happened shortly after the radiation--matter equality of Eq.~(\ref{eq:rm}) and long before the transition to vacuum--driven acceleration,  Eq.~(\ref{eq:lambdapoint}). Therefore we can ignore $\Lambda$ during the formation of the CMB and work only with $\Omega_\mathrm{R}$ and $\Omega_\mathrm{M}$. We define a rescaled scale factor $n$ and rescaled dimensionless time $\vartheta$ as
%%%%%%
\begin{equation}
a = \frac{\Omega_\mathrm{R}}{\Omega_\mathrm{M}}\, n \,,\quad 
\frac{\Omega_\mathrm{M}^2}{\Omega_\mathrm{R}^{3/2}}\,H_0 \,t = \vartheta \,.
\label{eq:rmscaling}
\end{equation}
%%%%%% 
We scale conformal time $\tau$ like cosmological time $t$ but use the same symbol for simplicity. We obtain from Eqs.~(\ref{eq:tauhubble}) and (\ref{eq:lambdacdm}):
%%%%%%
\begin{equation}
\tau = \int_0 \frac{\mathrm{d}n}{\sqrt{1+n}} = 2\sqrt{1+n}-2
\label{eq:taurm}
\end{equation}
%%%%%% 
where we put the zero of conformal time to $n=0$. We thus get the simple result:
%%%%%%
\begin{equation}
n = \tau +\frac{\tau^2}{4} \,.
\label{eq:nrmtau}
\end{equation}
%%%%%% 
We see that expansion and conformal time coincide initially, $n\sim\tau$ for $\tau\ll 1$. As light follows $\tau$ here it follows the expansion, and vice versa: the expansion lies on the light cone in the initial, radiation--dominated era. The initial cosmological horizon is light--like (Fig.~\ref{fig:horizons}). Consider now the expansion in cosmological time $\vartheta$. We obtain from Eq.~(\ref{eq:lambdacdm}):
%%%%%%
\begin{equation}
\vartheta = \int_0 \frac{n\,\mathrm{d}n}{\sqrt{1+n}} = - \frac{2}{3}(2-n)\sqrt{1+n} + \frac{4}{3} \,.
\label{eq:thetarelation}
\end{equation}
%%%%%% 
Solving Eq.~(\ref{eq:thetarelation}) for $n$ amounts to solving a cubic equation. We obtain from the trigonometric Cardano formula:
%%%%%%
\begin{equation}
n=\begin{cases}
1 + 2\cos\left(\frac{1}{3}\mathrm{arccos}\,\eta - \frac{2\pi}{3}\right) &: 0<\vartheta<\frac{4}{3} \\
1 + 2\cos\left(\frac{1}{3}\mathrm{arccos}\,\eta \right) &: \frac{4}{3}<\vartheta<\frac{8}{3} \\
1 + 2\cosh\left(\frac{1}{3}\mathrm{arcosh}\,\eta \right) &: \vartheta>\frac{8}{3} 
\end{cases} 
\end{equation}
%%%%%% 
with the abbreviation
%%%%%%
\begin{equation}
\eta = 1-3\vartheta+\frac{9}{2}\,\vartheta^2 \,.
\end{equation}
%%%%%% 
This is an analytic solution for the cosmic expansion in the radiation--matter era without $\Lambda$ and an excellent approximation of the full dynamics until $t_\mathrm{*}$ (with $\vartheta_\mathrm{*}=2.92$, $n_\mathrm{*}=3.17$). In terms of the rescaled scale factor the radiation--matter equilibrium occurs at $n=1$ and rescaled time
%%%%%%
\begin{equation}
\vartheta_\mathrm{eq} = \frac{4}{3}\left(1-2^{-1/2}\right) \approx 0.4 \,.
\end{equation}
%%%%%%
We obtain from our solution for the radiation--dominated era \cite{LL2}
%%%%%%
\begin{equation}
n\sim \sqrt{2\vartheta} \quad:\quad 0<\vartheta\ll \vartheta_\mathrm{eq} 
\end{equation}
%%%%%% 
and in the matter--dominated regime \cite{LL2}
%%%%%%
\begin{equation}
n\sim\left(3\vartheta/2\right)^{2/3} \quad\mbox:\quad \vartheta\gg \vartheta_\mathrm{eq} \,.
\label{eq:nm}
\end{equation}
%%%%%% 
However, with growing cosmic expansion matter domination will also come to an end, as $\Lambda$ takes over.

Now turn to the matter--vacuum period. During this era the radiation contribution to the cosmic dynamics becomes insignificant such that we can ignore $\Omega_\mathrm{R}$. It is wise to introduce a logarithmic scale for the scale factor, as $a$ grows by $10^3$ until the present (and is going to grow exponentially later on). We write $a$ as
%%%%%%
\begin{equation}
a = \left(\frac{\Omega_\mathrm{M}}{\Omega_\Lambda}\right)^{1/3} \mathrm{e}^\nu
\label{eq:anu}
\end{equation}
%%%%%% 
with logarithmic scale $\nu$ where at the present cosmological time ($a=1$)
%%%%%%
\begin{equation}
\nu_0 = \frac{1}{3} \,\ln \frac{\Omega_\Lambda}{\Omega_\mathrm{M}} \,,
\label{eq:nu0}
\end{equation}
%%%%%% 
and obtain from the $\Lambda$CDM model (\ref{eq:lambdacdm}):
%%%%%%
\begin{equation}
H_0 t \sqrt{\Omega_\Lambda} = \int \frac{\mathrm{d}\nu}{\sqrt{\mathrm{e}^{-3\nu}+1}}
= \frac{2}{3}\,\mathrm{arsinh}\left( \mathrm{e}^{3\nu/2}\right) 
\label{eq:lambdamint}
\end{equation}
%%%%%% 
with $t=0$ for $\nu\rightarrow-\infty$. Putting $\nu$ to the present, Eq.~(\ref{eq:nu0}), we obtain a good approximation for the cosmological age of the universe (as the radiation--dominated era is short in comparison with $t_0$):
%%%%%%
\begin{equation}
t_0 = \frac{1}{\sqrt{\Omega_\Lambda} H_0} \sinh\sqrt{\frac{\Omega_\Lambda}{\Omega_\mathrm{M}}} \,.
 \end{equation}
%%%%%% 
Finally, we obtain from Eqs.~(\ref{eq:anu}) and (\ref{eq:lambdamint}) the result:
%%%%%%
\begin{equation}
a = \left(\frac{\Omega_\mathrm{M}}{\Omega_\Lambda} \sinh ^2 x \right)^{1/3} \quad\mbox{with}\quad x = \frac{3}{2}\sqrt{\Omega_\Lambda} H_0\, t \,.
\end{equation}
%%%%%% 
For $H_0\,t\ll1$ we approximate $\sinh x\sim x$ and retrieve the matter--dominated regime of Eq.~(\ref{eq:nm}) with the scaling of Eq.~(\ref{eq:rmscaling}). For $H_0\,t\gg 1$ $\sinh$ approaches $\exp$ and thus the universe enters the phase of exponential expansion, dominated by $\Lambda$, its final phase.

We have determined the scalings and analytical solutions for the cosmic expansion in the radiation--matter era when the CMB was formed and in the matter--vacuum era when it freely propagates. Let us now conclude this section by working out how the cosmic parameters are retrieved from the correlations in the CMB. The correlation pattern (Fig.~\ref{fig:correlations}) follows the amplitudes of the acoustic waves as a function of wavenumber $k$ at the time of last scattering $t_\mathrm{*}$ (Fig.~\ref{fig:waves}). The details of these waves and how they contribute to the CMB are explained in Appendix A. Important for our argument here is only their phase $\varphi_\mathrm{*}$ at the time of last scattering. The phase is the most precisely measurable aspect of a wave. The phase is given by the wavenumber $k$ times the integral of the wave velocity over the propagation time. In our case, the wave velocity is the speed of sound $c_\mathrm{s}$ with respect to conformal time (if we integrate over conformal time):
%%%%%%
\begin{equation}
\varphi_\mathrm{*} = k \int_0^{\tau_\mathrm{*}} c_\mathrm{s}\, \mathrm{d}\tau = k \int_0^{n_\mathrm{*}} \frac{c_\mathrm{s}\,\mathrm{d} n}{\sqrt{1+n}}
\label{eq:phase}
\end{equation}
%%%%%% 
where we use scaled quantities according to Eq.~(\ref{eq:rmscaling}) and, in the last step, applied Eq.~(\ref{eq:taurm}). Note that we have written the wavenumber $k$ in inverse space units with space measured as $c$ times the scaled time $\vartheta$. In the CMB literature the phase integral $\varphi_\mathrm{*}/k$ is called the sound horizon. It corresponds to the distance acoustic information can travel from the initial time to the time of last scattering. 

Now, the phase (\ref{eq:phase}) depends on the scale factor $n_\mathrm{*}$ of last scattering. It explicitly depends on $n_\mathrm{*}$ though the integration limit and it depends on $n_\mathrm{*}$ implicitly: the speed of sound $c_\mathrm{s}$ varies with scale factor. In addition, $c_\mathrm{s}$ also depends on the fraction $\eta_\mathrm{B}$ of baryonic matter in $\Omega_\mathrm{M}$. The details (Appendix A) are not important here; important is the fact that the dependancies of $\varphi_\mathrm{*}$ on $n_\mathrm{s}$ and $\eta_\mathrm{B}$ are known. Important is also\footnote{The phase integral of Eq.~(\ref{eq:phase}) is not simply proportional to $n_\mathrm{*}$ due to the variation of $c_\mathrm{s}$ and the factor $(1+n)^{-1/2}$ caused by the cosmic expansion with Hubble parameter $H=n^{-2}(1+n)^{1/2}$.} that $\varphi_\mathrm{*}$ is not just proportional to $n_\mathrm{*}$. By fitting $\varphi_\mathrm{*}$ to the phase of the actual correlation curve (Fig.~\ref{fig:correlations}) one can thus retrieve $n_\mathrm{*}$ and $\eta_\mathrm{B}$ without prior knowledge of the scale of $k$. 

Known is also the physics of recombination \cite{Weinberg} predicting the recombination temperature $T_\mathrm{*}$ as a function of $\Omega_\mathrm{R}/\Omega_\mathrm{M}$. From the precisely measured average temperature $T_0$ of the CMB [Eq.~(\ref{eq:temperature0})] and the way (\ref{eq:temperature}) how temperature falls with scale factor one relates $a_\mathrm{*}$ to $T_0$ and $\Omega_\mathrm{R}/\Omega_\mathrm{M}$. Since $a_\mathrm{*}=(\Omega_\mathrm{R}/\Omega_\mathrm{M})\,n_\mathrm{*}$ and $n_\mathrm{*}$ is inferred from the CMB, one can thus obtain $\Omega_\mathrm{R}/\Omega_\mathrm{M}$. From the Stefan--Boltzmann law (\ref{eq:stefanboltzmann}) then follows $\Omega_\mathrm{M} H_0^2$. It remains to determine $\Omega_\Lambda H_0^2$, the vacuum contribution. 

After the release of the CMB the phase pattern $\varphi_\mathrm{*}$ is frozen and just gets rescaled by free electromagnetic propagation in the expanding universe. In conformal time $\tau$ this reduces to the propagation in non--expanding space. Hence we have for the distance $d$ the CMB has traveled:
%%%%%%
\begin{equation}
d = c \tau_\mathrm{*} \,,\quad \tau_\mathrm{*} = \int_{a_\mathrm{*}}^1 \frac{\mathrm{d} a}{a^2 H} \sim \int_0^1 \frac{\mathrm{d} a}{a^2 H}\,.
\label{eq:int}
\end{equation}
%%%%%% 
The free propagation of the CMB happens in the matter--$\Lambda$ era with $a$ given by Eq.~(\ref{eq:anu}) and $H$ by the relation $H^2=H_0^2 \Omega_\Lambda(\mathrm{e}^{3\nu}+1)$ in the $\Lambda$CDM model (\ref{eq:lambdacdm}) with $\Omega_\mathrm{R}=0$. Solving the integral (\ref{eq:int}) we obtain
%%%%%%
\begin{equation}
\tau_\mathrm{*} = \frac{2}{\sqrt{\Omega_\mathrm{M}} H_0}\,F(\Omega_\Lambda/\Omega_\mathrm {M})
\label{eq:tauL}
\end{equation}
%%%%%% 
with 
$
F(x)= {}_2 F_1\left(\frac{1}{6}, \frac{1}{2}, \frac{7}{6}; -x\right) 
$
in terms of Gauss' hypergeometric function. Equation~(\ref{eq:angle}) (Fig.~\ref{fig:angle}) relates the angle $\phi$ of a correlation in the CMB to the wavenumber $k$ (in scaled units). We get
%%%%%%
\begin{equation}
\phi = \left(\frac{\Omega_\mathrm{M}}{\Omega_\mathrm{R}}\right)^{3/2} \!\! \frac{\pi}{c k F(\Omega_\Lambda/\Omega_\mathrm {M})} \,.
\end{equation}
%%%%%% 
From the angular scale we can thus reconstruct the acoustic scale. As $F$ depends on the unknown $\Omega_\Lambda H_0^2$ divided by the known $\Omega_\mathrm{M} H_0^2$ we may retrieve $\Omega_\Lambda H_0^2$ and from Eq.~(\ref{eq:omegasum}) get $\Omega_\Lambda$ and $H_0$. Note that spatial curvature would distort and not simply scale the correlation curve \cite{Dodelson}. The apparent lack of distortion puts a bound on curvature \cite{CMBPlanck}. Note also that in our argument we have disentangled the retrieval of the cosmic parameters for conceptual simplicity. What counts in practice is maximal accuracy. In practice \cite{CMBPlanck} one should fit all parameters  together, and of course to the full correlation curve (Fig.~\ref{fig:correlations}) and not only to the phase. The most accurate quantity is the acoustic scale of the correlation pattern that gives us the Hubble constant $H_0$.

\section{Hubble tension}

The Hubble constant $H_0$ of Eq.~(\ref{eq:parameters}) has been inferred from the Cosmic Microwave Background, but it can also be measured directly from Hubble's law (\ref{eq:hubblelaw}) that relates redshift $-\mathrm{d}\omega/\omega$ to distance $\mathrm{d}\ell$. Frequencies $\omega$ and their shifts $\mathrm{d}\omega$ are measured precisely by spectroscopy. Astronomical distances are determined by distance ladders \cite{Riess22}. The first ladder is constructed from geometrical measurements taking the orbit of Earth around the Sun as baseline. Parallaxes (angle deviations) of stars are observed when Earth moves and the distances to those stars determined by triangulation. This first ladder reaches to about $10\mathrm{Kpc}$. The second ladder makes use of bright, pulsating stars --- Cepheids with a robust relation between brightness and pulsation rate \cite{Leavitt}. The apparent brightness depends on the actual brightness divided by the luminosity distance squared. The Cepheid relation is calibrated by geometrical measurements for the ones in sufficiently close proximity. Then, in turn, the established relation is used for inferring distances from the observed brightnesses, which constitutes the second ladder. This ladder reaches up to $40\mathrm{Mpc}$. The third ladder uses type Ia supernova explosions as ``standard candles''. These explosions are sufficiently bright to be measured up to redshifts $z\sim3$ and their actual brightness does not vary much (with deviations corrected from features of their spectra \cite{Riess22}). Supernovae are rare events (1 per century in an average galaxy) but 19 of them have been calibrated with Cepheids \cite{Riess22}. Having calibrated some type Ia supernovae as standard candles, many more similar supernovae serve to determine the distance as a function of redshift $z$ (Fig.~\ref{fig:hubble}). Performing the limit $z\rightarrow 0$ in the data gives $H$ at the present time, with nearly $1\%$ precision \cite{Riess22}:
%%%%%%
\begin{equation}
H_0 = (73.04 \pm 1.04) \,\frac{\mathrm{km}/\mathrm{s}}{\mathrm{Mpc}} \,.
\label{eq:hubblemeasured}
\end{equation}
%%%%%% 
The problem is that this value does not agree with the $H_0$ in the cosmic parameters (\ref{eq:parameters}) inferred from the CMB. The uncertainty in the measured value (\ref{eq:hubblemeasured}) has been reduced to $5\sigma$ \cite{Riess22}, which in high--energy physics would qualify as a discovery. In cosmology, there are still sceptics \cite{Efstathiou} but the tension between $H_0$ inferred from the early universe (from the CMB) and the Hubble constant measured by late--universe probes persists also in other observations based on completely different methods \cite{DiValentino} although with less precision yet. The overall evidence \cite{DiValentino} indicates that the Hubble tension is real. 

There are about $10^2$ theories \cite{DiValentinoTheories} explaining the Hubble tension, but all of them require untested modifications to the standard model of particle physics or to general relativity or to the cosmological principle, except one \cite{Berechya}. This theory is based \cite{Annals} on the physics of the quantum vacuum \cite{Milonni,Milton,KMM,BKMM,Rodriguez,Lambrecht,Buhmann,Forces}. It does make extrapolations  --- it extrapolates quantum electrodynamics from dielectric media to curved space--time \cite{LeoPhil} --- and it contains assumptions on renormalization \cite{Annals,Grin1}, but its principal steps are empirically tested \cite{KMM,Rodriguez,Lambrecht} or experimentally testable \cite{EPL,Grin2,Efrat}. 

Here is the argument \cite{Berechya} of how the Casimir effect explains the Hubble tension. We have seen in Sec.~4 that $H_0$ is inferred from the relation between the angular and the acoustic scale of the CMB. This relation requires knowing the distance $d$ to the surface of last scattering (Fig.~\ref{fig:angle}) and $d$ is given by Eq.~(\ref{eq:int}). It is the only quantity that depends significantly on the cosmological constant $\Lambda$, because only during the free propagation of the CMB in the matter--vacuum era $\Lambda$ plays a significant role. The shape of the correlation curve (Fig.~\ref{fig:correlations}) is formed in the radiation--matter era when the CMB is formed (Fig.~\ref{fig:waves}). So the shape of the curve depends on $\Omega_\mathrm{R}$ and $\Omega_\mathrm{M}$ (and on the contributions from neutrinos and dark matter) but not on $\Omega_\Lambda H_0^2$, only the scale of the curve does. 

Now, the $\Lambda$CDM model assumes that the cosmological constant $\Lambda$ is constant. Where is the evidence for this? The CMB data do not prove that $\Lambda$ is constant, because $\Lambda$ is contained in only one parameter, the scale of the correlation curve. As long as $\Lambda$ is concerned, all the CMB data condense into one point. A single data point does not prove constancy. The Hubble diagram (Fig.~\ref{fig:hubble}) maps out the late stage of the cosmic expansion as a curve depending on the constancy of $\Lambda$. A constant $\Lambda$ is consistent with the data, so $\Lambda$ must not vary much, but the observed $H_0$ is inconsistent \cite{Riess22}. Is there good theoretical evidence? The most--common and most--convincing argument is this: the vacuum should be generally covariant --- it should look the same in all reference frames. Consequently, the energy--momentum tensor \cite{LL2} of the vacuum must be a constant, the cosmological constant. This argument is wrong, because the premise is wrong: the vacuum is not generally covariant. For example, in the Unruh--Fulling--Davis effect the vacuum state of Minkowski space appears as thermal radiation to an accelerated observer. The vacuum is frame--dependent, as if it were a physical substance. The vacuum is an unusual ``substance'' though --- it does not resist uniform motion, but it resists acceleration. In any case, the vacuum is not generally covariant, which invalidates the theoretical argument for the constancy of $\Lambda$. Can a constant $\Lambda$ be calculated from theory? Zel'dovich suggested that $\Lambda$ is given by the vacuum energy and calculated it for a simple model \cite{Zeldovich}. However, this and other calculations disagree with the actual value of $\Lambda$ by many orders of magnitude, depending on the cut--off of the theory \cite{WeinbergLambda}. There is therefore no credible evidence for the absolute constancy of $\Lambda$. 

Suppose now that $\Lambda$ is given by the vacuum energy as obtained in Casimir physics \cite{Milonni,Milton,KMM,BKMM,Rodriguez,Lambrecht,Buhmann,Forces}. We can apply this theory, because \cite{LeoPhil} curved space--time appears as a dielectric medium to electromagnetic fields \cite{Gordon,Quan1,Quan2,Plebanski,Schleich,Stor,GREE}. The many orders of magnitude disagreement with data disappear by renormalization \cite{Annals}. What remains is an energy density $\varepsilon_\mathrm{vac}$ of the correct order of magnitude \cite{Annals}. Why this is so is subtle and is discussed elsewhere \cite{Annals,London}. Here we describe how $\varepsilon_\mathrm{vac}$ influences the cosmic dynamics \cite{Berechya} and resolves the Hubble tension \cite{DiValentino,Riess22} using a thermodynamic argument \cite{Annals}.

In the standard Casimir effect, the vacuum exerts forces on the interfaces of dielectric media. Spatially varying media create spatially varying vacuum energies and stresses. In cosmology, space is homogeneous, but expands in time $t$. The varying scale factor $a(t)$ creates a non--trivial vacuum energy \cite{Annals} similar to the dynamical Casimir effect \cite{SchwingerDCE,Mendonca1,Mendonca2,Dodonov,Wilson,Hakonen,Veccoli} briefly discussed in Sec.~2. Since $\varepsilon_\mathrm{vac}$ is created by variations in $a$ it must depend on derivates of $a$. But this implies that $\varepsilon_\mathrm{vac}$ is not like the other energy densities, $\varepsilon_\mathrm{R}$ and $\varepsilon_\mathrm{M}$, that only depend on $a$. They follow the cosmic expansion adiabatically, whereas $\varepsilon_\mathrm{vac}$ is non--adiabatic, which violates the thermodynamic relation (\ref{eq:f2}). This relation is necessary \cite{LL2} for the energy--momentum conservation and the validity of Einstein's field equations of gravity \cite{LL2}. For maintaining the energy balance we thus need to add an energy density $\varepsilon_\Lambda$ to $\varepsilon_\mathrm{vac}$. For the vacuum pressure $p_\mathrm{vac}$ we have
%%%%%%
\begin{equation}
p_\mathrm{vac}=\frac{\varepsilon_\mathrm{vac}}{3} 
\label{eq:vacuumpressure}
\end{equation}
%%%%%% 
like in Eq.~(\ref{eq:radiationpressure}) for any other incoherent radiation pressure. The additional $\varepsilon_\Lambda$ should just take care of the energy balance in Eq.~(\ref{eq:f2}), but not cause  an imbalance itself. We thus require $p_\Lambda=-\varepsilon_\Lambda$ [Eq.~(\ref{eq:plambda})] and obtain from Eq.~(\ref{eq:f2}):
%%%%%%
\begin{equation}
\partial_t(\varepsilon_\Lambda +\varepsilon_\mathrm{vac}) = -4H\varepsilon_\mathrm{vac} \,.
\label{eq:lambdadynamics}
\end{equation}
%%%%%% 
The additional $\varepsilon_\Lambda$ with pressure $-\varepsilon_\Lambda$ appears like the trace anomaly\footnote{For a physical interpretation of the trace anomaly see Refs.~\cite{London} and \cite{Efrat}.} of conformally invariant fields \cite{Wald}. It shows the characteristic feature, $p_\Lambda=-\varepsilon_\Lambda$, of the cosmological constant, but $\varepsilon_\Lambda$ varies. However, the dynamical equation (\ref{eq:lambdadynamics}) allows for an integration constant we may put to
%%%%%%
\begin{equation}
\varepsilon_\infty = \lim_{a\rightarrow\infty} \varepsilon_\Lambda
\label{eq:lambdainfty}
\end{equation}
%%%%%% 
and interpret as the cosmological constant at infinity when space is expanding exponentially with constant Hubble parameter $H$. The vacuum energy $\varepsilon_\mathrm{vac}$ is generated \cite{Annals} by variations in the Gibbons--Hawking temperature of Eq.~(\ref{eq:gh}). In the era of exponential expansion, these variations vanish and hence $\varepsilon_\mathrm{vac}\rightarrow 0$, only $\varepsilon_\Lambda$ remains with $\varepsilon_\Lambda\rightarrow\varepsilon_\infty$. 

The vacuum--dominated era is thus characterized by a constant cosmological constant. However, in the transition period between matter and vacuum domination --- the period we live in --- $\varepsilon_\mathrm{vac}$ varies and so does $\varepsilon_\mathrm{\Lambda}$. The variation in the sum of the two energies affects the cosmic dynamics and hence the distance $d$ required for relating the angular scale of the CMB to the acoustic scale (Fig.~\ref{fig:angle}). The $H_0$ inferred from that distance assuming constant $\varepsilon_\Lambda$ is no longer the actual Hubble constant, because $\varepsilon_\Lambda$ has varied. We may adjust the integration constant $\varepsilon_\infty$ such that we fit the angular scale of the CMB \cite{Berechya} and then see what the dynamics of Eq.~(\ref{eq:lambdadynamics}) does. For this we need to determine $\varepsilon_\mathrm{vac}$ from Casimir physics \cite{Annals,Berechya} assuming a cut--off close to the scale where space--time presumably ceases to act like a medium, the Planck scale characterized in terms of the Planck length 
%%%%%%
\begin{equation}
\ell_\mathrm{P}= \sqrt{\frac{\hbar G}{c^3}} \approx 10^{-35} \mathrm{m} \,. 
\label{eq:plancklength}
\end{equation}
%%%%%% 
It turns out \cite{Berechya} that, for a cut--off at exactly the Planck length and for the most naive model, the theoretical variation of the Hubble constant agrees with the latest data [8.4\%, Eq,~(\ref{eq:hubblemeasured})] with $1\%$ precision, the very precision of the data \cite{Riess22}.

\section{Outlook}

Casimir physics has a chance of resolving the biggest crisis in contemporary astrophysics. The first results \cite{Annals,Berechya} seem promising, but much more work is needed to be certain that they are real and not coincidences. 
{\it First}, the theory developed so far \cite{Annals} is incomplete: While the theory takes the Gibbons--Hawking effect of cosmological horizons (Sec.~2) as its starting point, it ignores the Hawking partners created on the other side of the horizon. As the cosmological horizon is not an event horizon (Fig.~\ref{fig:horizons}) the partners eventually arrive and play a role of equal importance than the Hawking particles created this side of the horizon. 
{\it Second}, quantum electromagnetism and quantum electromagnetism alone seems to fit the bill. Where are the other fields of the standard model of particle physics, why are they silent here? 
{\it Third}, further and more consistent analysis of the astronomical data is needed. A great deal of modern cosmology is statistics: exploring variations in the parameters and their influences on several, disconnected data sets from vastly different cosmological eras (CMB, Baryon Acoustic Oscillations, supernovae, gravitational lensing etc). 
{\it Fourth}, there are other tensions in the cosmological data than the Hubble tension, for example inconsistencies in the standard fit to the CMB correlations (Fig.~\ref{fig:correlations}) for low angular Fourier components $l$ or inconsistencies in the value of the matter density and the amplitude of cosmic structure \cite{Tensions}. These tensions are more subtle and less statistically significant ($\sim 3\sigma$ instead of $5\sigma$ for the Hubble tension \cite{Riess22}) but they might be real, too. If Casimir physics resolves the Hubble tension it might also resolve the other tensions \cite{Tensions}. 
{\it Fifth}, experimental tests are needed. These should be tests of the principle mechanism and could be performed in laboratory analogues of gravity. For example, the anomaly giving rise to the cosmological constant \cite{Annals} can be tested with ultracold atoms \cite{Efrat}. Or the connection \cite{EPL} between the Gibbons--Hawking effect and the dynamical Casimir effect can be explored in the laboratory. There are probably more (and better) tests possible. They could give much needed empirical grounding to the lofty extrapolations of the theory, which brings us finally to: 
{\it Sixth}, the foundations of the theory need to be better understood. While the Lifshitz theory of the cosmological constant \cite{Annals} is based on well--understood and well--tested physics, it still makes extrapolations and assumptions. In particular, the renormalization method of the theory needs a better foundation. The problem is related to the renormalization of the Casimir effect in inhomogeneous media. In cosmology, space represents a uniform medium inhomogeneous in time, on Earth graded--index materials are constant in time but inhomogeneous in space. While the Casimir effect in inhomogeneous media has been full of surprises \cite{Simpson} the theory has remained underdeveloped since the foundation of the field by Lifshitz, Dzyaloshinskii and Pitaevskii \cite{Lifshitz,LDP}. There was simply no need, no urgent problem to be solved in inhomogeneous media --- and the theory is difficult and observable effects, if any, are likely to be obscure. So is it surprising that inhomogeneous media have been a backwater? Now there is an urgent problem worth the salt of the Casimir community. Let's get it solved. 

\section*{Acknowledgements}
I am most grateful to Dror Berechya for his helpful comments and the many discussions we had during our struggle to understand modern cosmology, and to
David Bermudez and Nikolay Ebel for discussions and comments on analogues of gravity and beyond. 
The paper has been supported by the Israel Science Foundation and the Murray B. Koffler Professorial Chair. 

\appendix

\renewcommand{\theequation}{A\arabic{equation}}
\setcounter{equation}{0}

\section{Appendix: Cosmic fluctuations}

In this appendix we develop the fluid--mechanical theory of cosmic fluctuations. Such fluctuations are the ones observed in the Cosmic Microwave Background (CMB). They are also responsible for the onset of structure formation --- they are the seeds of gas clouds, galaxies, stars, us. The accurate theory of cosmic fluctuation, capable of fully fitting the observed correlation spectrum (Fig.~\ref{fig:correlations}) requires the use of relativistic Boltzmann equations \cite{Dodelson} and of computer simulations. However, ``it is nice to know that the computer understands the problem, but I want to understand it, too'' (Wheeler). In this appendix we deduce simple analytic approximations for the acoustic waves of light and matter that create the features of the cosmic fluctuations, including the undulations in the correlation curve (Fig.~\ref{fig:correlations}). The results require simplifications of the physical assumptions and are therefore not perfectly accurate, but they qualitatively describe the essentials with just elementary functions. Such a programme of fluid--mechanical approximation had been attempted  before \cite{Weinberg} but still needed cumbersome connection procedures, here we get it done. 

\subsection{Setting the scene}

We will follow the same logic as in the main part of this article. We begin by stating the space--time geometry. Then we describe the content of the universe by fluids. We work out a thermodynamical relation for the fluctuations connecting number density with energy/mass density. We deduce the consequences of number and energy--momentum conservation, and of gravity. In the regime of weak fluctuations and of weak gravity we condense the fully relativistic dynamics in minimal modifications of the non--relativistic laws \cite{Falkovich}: the equation of continuity for the number density, the Bernoulli equation for the energy density, and the Poisson equation for gravity. We then solve these equations analytically for matter coupled to radiation.

Relativity is needed, because we need to deal with relativistic fluids such as light and neutrinos made of particles moving at (or close to) the speed of light $c$. For keeping the notation uncluttered, we put in our calculations 
%%%%%%
\begin{equation}
c = 1 \,,
\end{equation}
%%%%%% 
which simply means we count time by $ct$ (or measure space by light). In our results we will point out where $c$ and hence relativity enters. We describe the space--time geometry in comoving coordinates by the metric: 
%%%%%%
\begin{equation}
\mathrm{d}s^2 =  (1+2\Phi)\,\mathrm{d}t^2 - a^2 (1-2\Phi)\,\mathrm{d}\bm{r}^2 
\label{eq:metric1}
\end{equation}
%%%%%% 
where as before the scale factor $a$ is a function of cosmological time $t$, but now we also consider the gravitational potential $\Phi$ that may vary in space and time. The metric (\ref{eq:metric1}) is the perturbed Friedman--Lemaitre--Robertson--Walker metric (\ref{eq:metric}) in Newtonian gauge. We assume weak gravity:
 %%%%%%
\begin{equation}
|\Phi|\ll 1 \,.
\label{eq:weak}
\end{equation}
%%%%%% 
The gravitational potential has the dimensions of a velocity squared; Eq.~(\ref{eq:weak}) means that the magnitude of $\Phi$ is much smaller than $c^2$. The term $(1+2\Phi)\,\mathrm{d}t^2$ in Eq.~(\ref{eq:metric1}) describes the gravitational redshift, the change in the measure of time $\sqrt{g_{00}}\sim1+\Phi$, connecting Einsteinian space--time geometry to Newtonian gravity. The term $a^2(1-2\Phi)\,\mathrm{d}\bm{r}^2$ accounts for the scale factor $a$ due to cosmic expansion and the spatial contraction gravity causes. The factor $(1-2\Phi)$ is only relevant for relativistic velocities and relativistic fluids such as light, but there it is essential. For example, without the spatial factor, a gravitational lens \cite{Schneider} would deflect light only by half the correct angle, the other half is due to it. In the metric (\ref{eq:metric1}) the scale factor $a$ describes the overall expansion of space $|\mathrm{d}\bm{r}|$. The expanding universe sets the background of the scene. Note that although these considerations make the metric (\ref{eq:metric1}) plausible, we still need to prove its validity by putting it in Einstein's equations \cite{LL2}. In accordance with the condition (\ref{eq:weak})  of weak gravity, we will do this to linear order (Sec.~A.3).

The cosmic fluids (electromagnetic and neutrino radiation, baryonic and dark matter) we describe by their mass/energy densities $\varepsilon$ and pressures $p$. For notational simplicity, we focus on only one of the fluids. We combine $\varepsilon$ and $p$ in the enthalpy density:
 %%%%%%
\begin{equation}
w = \varepsilon + p \,.
\label{eq:enthalpy}
\end{equation}
%%%%%% 
In thermodynamics \cite{LL5}, the enthalpy is defined as $H=E+pV$ and $w$ denotes the density $H/V$. The central quantity of relativistic fluid mechanics \cite{LL6} is the energy--momentum tensor:
 %%%%%%
\begin{equation}
T^{\alpha\beta} = w\, u^\alpha u^\beta - p\,g^{\alpha\beta} 
\label{eq:tensor}
\end{equation}
%%%%%%
where $g_{\alpha\beta}$ is the metric tensor and $g^{\alpha\beta}$ its matrix--inverse, while $u^\alpha$ denotes the four velocity $\mathrm{d}x^\alpha/\mathrm{d} s$, satisfying $u_\alpha u^\alpha=1$. Throughout, we adopt Einstein's summation convention. Whenever we use index notation, Greek indices refer to the space--time coordinates $\{t,\bm{r}\}$ and run from $0$ to $3$, while Latin indices refer to the spatial coordinates running from $1$ to $3$. The energy--momentum tensor (\ref{eq:tensor}) is justified as follows. In a Galilean frame \cite{LL2} locally comoving with the fluid $u^\alpha=(1,\bm{0})$ and so the only non--vanishing components of $T^{\alpha\beta}$ are $T^{00}=\varepsilon$ and $T^{ii}=p$ as required. Since expression (\ref{eq:tensor}) is covariant it must also hold in any frame, which proves Eq.~(\ref{eq:tensor}). 

In fluid mechanics \cite{LL6,Falkovich}, small oscillations such as sound waves correspond to potential flows with $\mathrm{curl}\,\bm{v}=0$ for the three--dimensional velocity. We therefore express $\bm{v}$ in terms of the velocity potential $\varphi$ \cite{Falkovich} as $v_i=-\partial_i \varphi$ for covariant $v_i$ and hence for the contravariant $\bm{v}=v^i$ as:
%%%%%%
\begin{equation}
\bm{v} = \frac{\nabla\varphi}{a^2}
\label{eq:velocitypotential}
\end{equation}
%%%%%%
where we have assumed small velocity perturbations in comparison with $c$ and weak gravity such that the metric matters only to leading order. Note that the velocity perturbations are small, even for light. On average light fills space uniformly with zero mean velocity, and the deviations $v$ are small. In this regime we retrieve from $u_\alpha u^\alpha=1$ the $u^0$ component; we include it in the expression
%%%%%%
\begin{equation}
u^\alpha = \left(1-\Phi, \bm{v}\right) 
\label{eq:fourvelocity}
\end{equation}
%%%%%%
to linear order. We also linearize the energy/mass density and pressure around the average energy/mass density $\bar{\varepsilon}$ we denote as $\rho$ and the average pressure $\bar{p}$ we denote in terms of the average enthalpy density $W$ (for better distinguishing $\bar{\varrho}$ and $\bar{p}$ from $\varrho$ and $p$):
%%%%%%
\begin{equation}
\varepsilon = (1+\delta)\rho \,,\quad p = W-\rho +c_\mathrm{s}^2 \delta\rho
\label{eq:densitypressure}
\end{equation}
%%%%%%
where $\delta$ describes the density deviation, also called the over density, and $c_\mathrm{s}$ is defined by
%%%%%%
\begin{equation}
c_\mathrm{s}^2 = \frac{\partial \bar{p}}{\partial\rho} = \frac{\partial (W-\rho)}{\partial\rho} \,.
\label{eq:defsound}
\end{equation}
%%%%%%
We will show that $c_\mathrm{s}$ is the speed of sound (with respect to conformal time). For radiation and matter we get from Eqs.~(\ref{eq:radiationpressure}) and (\ref{eq:matterpressure}):
%%%%%%
\begin{equation}
W_\mathrm{R} =\frac{4}{3}\,\rho_\mathrm{R} \,,\quad W_\mathrm{M}=\rho_\mathrm{M} 
\label{eq:radmat}
\end{equation}
%%%%%%
and hence, re--instituting $c$,  the speed of sound $c_\mathrm{R}=c/\sqrt{3}$ for radiation and zero sound speed (in comparison with $c$) for matter. 

In addition to energy and pressure, we also need to consider the particle density we denote as $\varrho$ here to distinguish it from the mass density $\rho$. The average number density in the expanding universe must go as $a^{-3}$. We describe deviations with the excess $\nu$ defined by
%%%%%%
\begin{equation}
\varrho = \frac{1+\nu}{a^3} \,.
\label{eq:particles}
\end{equation}
%%%%%%
We assume adiabatic fluids where the entropy is conserved. In this case, thermodynamics \cite{LL5} requires a relationship between density and particle fluctuations:
%%%%%%
\begin{equation}
\delta\rho = W\nu \,.
\label{eq:relation}
\end{equation}
%%%%%%
To derive this relation, consider $\mathrm{d}(\varepsilon V)=-p\,\mathrm{d} V$ with $V=\varrho^{-1}$ for Eqs.~(\ref{eq:densitypressure}) and (\ref{eq:particles}) and linearize. For spatial deviations the average densities do not vary, and we get $\nabla(\delta\rho-W\nu)=0$. For variation in time we obtain the thermodynamic relation (\ref{eq:f2}) in the form 
%%%%%%
\begin{equation}
\partial_t \rho = -3HW
\label{eq:f2w}
\end{equation}
%%%%%%
in zeroth order, with $H$ being the Hubble parameter. One then verifies that Eq.~(\ref{eq:relation}) satisfies $\partial_t (\varepsilon V)=-p\partial_t V$ to linear order as well, which concludes the proof. Equation (\ref{eq:relation}) relates the deviation in energy to the enthalpy carried by the particles of the fluid. The enthalpy $E+pV$ describes the energy including the effect of the pressure and is widely used in chemistry. Particles contribute with the enthalpy to the energy density. 

\subsection{Relativistic fluid mechanics}

In this subsection we derive the fundamental equations of motion for the fluid, the continuity equation \cite{Falkovich} for the excess $\nu$ in particle density and the Bernoulli equation \cite{Falkovich} for the density deviation, the velocity potential $\varphi$ and the gravitational potential $\Phi$. Let us first state the results and discuss them. We are going to get for the equation of continuity:
%%%%%%
\begin{equation}
\partial_t (\nu-3\Phi) + \frac{\nabla^2\varphi}{a^2} = 0 \,.
\label{eq:continuity}
\end{equation}
%%%%%%
This equation  accounts for the conservation of particle number. For this we need to take the expansion of the universe into account, which is included in the definition (\ref{eq:particles}) of $\nu$. The term $\nabla^2\varphi/a^2$ describes the Laplacian in a spatial geometry of line element $\mathrm{d}\ell=a\,|\mathrm{d}\bm{r}|$. Equation~(\ref{eq:continuity}) is nothing but the ordinary equation of continuity for a potential flow, apart from the $-3\Phi$ inside the time derivative. This term accounts for the fact that the spatial volume element changes not only due to expansion, but also due to gravity. The spatial volume element is given by the space--time volume element $\sqrt{-g}$ divided by the measure of time $\sqrt{g_{00}}$ \cite{LL2} where $g$ denotes the determinant of the metric tensor $g_{\alpha\beta}$. For the space--time volume element we get to linear order:
%%%%%%
\begin{equation}
\sqrt{-g} = a^3(1-2\Phi) \,.
\label{eq:g}
\end{equation}
%%%%%%
Hence the spatial volume element is $a^3(1-3\Phi)$ to linear order, which produces the correction $-3\Phi$ in the actual number density that appears in the equation (\ref{eq:continuity}) of continuity. From the general equation of continuity \cite{LL2} $\partial_\alpha \sqrt{-g} (\varrho u^\alpha) = 0$, and from expressions (\ref{eq:velocitypotential}) and (\ref{eq:fourvelocity}) and definition (\ref{eq:particles}) follows Eq.~(\ref{eq:continuity}) to linear order.

For a potential flow \cite{Falkovich} the Bernoulli equation describes the energy conservation along a stream line. We obtain in our case:
%%%%%%
\begin{equation}
W(\dot{\varphi} + \Phi) + c_\mathrm{s}^2(\delta\rho+\dot{\rho}\,\varphi) = 0 
\label{eq:bernoulli0}
\end{equation}
%%%%%%
where the dots denote time derivatives. In non--relativistic fluid mechanics \cite{Falkovich} the energy per mass on a stream line is given by $-\dot{\varphi}$. We see that in relativistic fluid mechanics the enthalpy plays the role of the mass. We also see that the equivalent $W$ of the inertial mass in front of $-\dot{\varphi}$ is the same as the $W$ the gravitational potential is multiplied with, which expresses the equivalence principle \cite{LL2} in our case. The term $c_\mathrm{s}^2(\delta\rho+\dot{\rho}\,\varphi)$ describes the internal pressure of the fluid. Here the density deviation gets a relativistic correction of $\dot{\rho}\,\varphi$. We remove the mass factor in the Bernoulli equation (\ref{eq:bernoulli0}) using the thermodynamic relations (\ref{eq:relation}) and (\ref{eq:f2w}):
%%%%%%
\begin{equation}
\partial_t \varphi + \Phi + c_\mathrm{s}^2(\nu -3H\varphi) = 0 
\label{eq:bernoulli}
\end{equation}
%%%%%%
with Hubble parameter $H$. The Bernoulli equation (\ref{eq:bernoulli}) tells how the pressure and the gravitational potential $\Phi$ sets perturbations of the fluid into motion. These pressure waves are sound waves. To see this, ignore gravity ($\Phi=0$) and expansion ($a=\mathrm{const}$) for a moment. Differentiate the Bernoulli equation (\ref{eq:bernoulli}) with respect to time and apply the equation (\ref{eq:continuity}) of continuity.  One arrives at the wave equation of sound: $\partial_t^2\varphi=(c_\mathrm{s}/a)^2\nabla^2\varphi$. Like for light, the scale factor plays the role of the refractive index. In conformal time (\ref{eq:tau})  $c_\mathrm{s}/a$ reduces to $c_\mathrm{s}$. We thus conclude that $c_\mathrm{s}$ describes the speed of sound with respect to conformal time. Unlike light, however, sound waves are not conformally invariant: when $a$ varies they cannot be fully reduced to plane waves with respect to conformal time (even when $\Phi=0$). 

In the plasma generating the CMB, light and (baryonic) matter are strongly coupled. They form {\it one} fluid of particle density $\varrho$ with excess $\nu$. Light and (baryonic) matter move with the same velocity and hence share the velocity potential $\varphi$. However, only light contributes to the pressure. In the original Bernoulli equation (\ref{eq:bernoulli0}) we need to replace $W$ by $W_\mathrm{R}+W_\mathrm{M}$ and $c_\mathrm{s}^2(\delta\rho+\dot{\rho}\,\varphi)$ by $c_\mathrm{R}^2(\delta\rho_\mathrm{R}+\dot{\rho}_\mathrm{R}\,\varphi)$. We arrive at the Bernoulli equation (\ref{eq:bernoulli}) with the effective speed of sound:
%%%%%%
\begin{equation}
c_\mathrm{s} = \frac{c_\mathrm{R}}{\sqrt{1+W_\mathrm{M}/W_\mathrm{R}}}\,.
\label{eq:sound}
\end{equation}
%%%%%%
We see that $c_\mathrm{s}$ depends on the relative enthalpies $W_\mathrm{M}/W_\mathrm{R}$. The enthalpies vary with the densities $\rho_\mathrm{R}$ and $\rho_\mathrm{M}$ according to Eq.~(\ref{eq:radmat}) and so the sound speed varies with density. In the radiation--dominated era just after the big bang, $W_\mathrm{M}/W_\mathrm{R}$ vanishes, so the sound speed of light amounts to $1/\sqrt{3}$ the speed of light $c$. In the matter--dominated era, the speed of sound vanishes, as only light exerts pressure. The CMB is formed in the transition period between radiation and matter domination and hence it contains information about their relative weight $\Omega_\mathrm{R}/\Omega_\mathrm{M}$. In particular, the phase of correlations in the CMB is given by the phase of acoustic waves, Eq.~(\ref{eq:phase}). The $c_\mathrm{s}$ that enters there is the effective speed of sound we have determined\footnote{However, only the baryonic part of matter and only the electromagnetic part of radiation contributes to the ratio $W_\mathrm{M}/W_\mathrm{R}$ in the actual sound speed, see Eq.~(\ref{eq:phasen}).} in Eq.~(\ref{eq:sound}).

It remains to derive the Bernoulli equation (\ref{eq:bernoulli0}) from relativistic fluid mechanics. In the first step we follow the standard procedure \cite{LL6} for a general relativistic fluid (with potential flow). We perform the analysis in a local geodesic frame \cite{LL2} where we are (locally) in Minkowski space. In such a frame the conservation of energy and momentum corresponds to $\partial_\alpha T^{\alpha\beta}=0$. Consider the projection $u_\beta \partial_\alpha T^{\alpha\beta}$. We obtain from the structure (\ref{eq:tensor}) of the energy--momentum tensor and $u_\beta u^\beta=1$ the relation $\partial_\alpha w u^\alpha=u^\beta\partial_\beta p$. In a local geodesic frame the continuity equation reads $\partial_\alpha \varrho u^\alpha=0$. Consequently,
%%%%%%
\begin{equation}
u^\alpha\left(\partial_\alpha\frac{w}{\varrho}- \frac{1}{\varrho}\,\partial_\alpha p\right) = 0\,.
\label{eq:isentropy}
\end{equation}
%%%%%%
This relation is automatically satisfied thanks to the thermodynamics of isentropic fluids where $\mathrm{d}(\varepsilon V)=-p\,\mathrm{d}V$ for $V=1/\varrho$ and $w=\varepsilon+p$. In the direction of the four--velocity, energy--momentum conservation reduces to thermodynamics. 

Now turn to the orthogonal components $\partial_\alpha T^\alpha_\mu-u_\mu u_\beta\partial_\alpha T^{\alpha\beta}$. For them we obtain from Eq.~(\ref{eq:tensor})  the expression $wu^\alpha\partial_\alpha u_\mu-\partial_\mu p - u_\mu u^\alpha\partial_\alpha p$ and finally from the adiabaticity condition (\ref{eq:isentropy}):
%%%%%%
\begin{equation}
u^\alpha \partial_\alpha \frac{w}{\varrho} \,u_\beta = \partial_\beta \frac{w}{\varrho} \,.
\label{eq:euler}
\end{equation}
%%%%%%
This is the relativistic Euler equation \cite{Falkovich}. For a potential flow we solve the Euler equation assuming Khalatnikov's ansatz \cite{Khalatnikov}:
%%%%%%
\begin{equation}
\frac{w}{\varrho}\,u_\alpha = - \partial_\alpha\chi
\label{eq:khalatnikov}
\end{equation}
%%%%%%
where $\chi$ is the relativistic velocity potential. In this case, the left--hand side of the Euler equation (\ref{eq:euler}) becomes $u^\alpha\partial_\beta(w/\varrho) u_\alpha$ and (as $u^\alpha u_\alpha=1$) gives the right--hand side $\partial_\beta(w/\varrho)$. Khalatnikov's expression (\ref{eq:khalatnikov}) condenses the condition for a potential flow and the Bernoulli equation in one formula, as we see next.

Khalatnikov's formula (\ref{eq:khalatnikov}) is manifestly covariant and hence valid in all frames, not only in the local geodesic frame in which it was derived. We can immediately apply it to our setting of gravitational fields in the expanding universe. To zeroth order $w/\varrho=Wa^3$ and therefore $\chi_0=-\int Wa^3\,\mathrm{d}t$. For the spatial components we have $u_i=-\nabla\varphi$ to first order, which suggests:
%%%%%%
\begin{equation}
\chi_1 = W a^3 \varphi \,.
\end{equation}
%%%%%%
Armed with this result we return to the time component of Eq.~(\ref{eq:khalatnikov}) and analize it to first order. Differentiating $\chi_1$ with respect to time gives $a^3(\dot{W}\varphi+W\dot{\varphi}+3H\varphi)$. From the definition (\ref{eq:defsound}) of the speed of sound and the thermodynamic relation (\ref{eq:f2w}) we obtain $\dot{W}+3HW=c_\mathrm{s}^2\dot{\rho}$. In first order, the left--hand side of Eq.~(\ref{eq:khalatnikov}) gives $-a^3 (W\Phi+c_\mathrm{s}^2\delta\rho)$ where we have used Eqs.~(\ref{eq:enthalpy}), (\ref{eq:fourvelocity}), (\ref{eq:densitypressure}), (\ref{eq:particles}) and (\ref{eq:relation}). We have thus derived the Bernoulli equation (\ref{eq:bernoulli0}).

\subsection{Gravity}

The pressure of light and the inertia of matter (and light) make acoustic waves. These waves are reshaped by the expanding universe and the gravitational potential $\Phi$, as described in the equation of continuity (\ref{eq:continuity}) and the Bernoulli equation (\ref{eq:bernoulli}). Now we need to work out how the gravitational potential is generated. We also need to justify our starting point, the metric (\ref{eq:metric1}) that encodes both the expansion and the gravitational potential. For this we will show that the metric (\ref{eq:metric1}) and the dynamical equations (\ref{eq:continuity}) and (\ref{eq:bernoulli}) are consistent with Einstein's field equations \cite{LL2}. The only new information contained in Einsteinian gravity will be the almost--Newtonian Poisson equation:
%%%%%%
\begin{equation}
\frac{\nabla^2\Phi}{a^2} = 4\pi G \left(\delta\rho+\dot{\rho}\,\varphi\right) .
\label{eq:poisson}
\end{equation}
%%%%%%
This equation describes how deviations from the average density and from the comoving expansion generate the gravitational potential. The left--hand side consists of the ordinary Laplacian of $\Phi$ in the universe of expanding length scale $a$. On the right--hand side we recognize the density fluctuations $\delta\rho+\dot{\rho}\,\varphi$ from the Bernoulli equation (\ref{eq:bernoulli0}). We use the thermodynamic relations (\ref{eq:relation}) and (\ref{eq:f2w}) to write the Poisson equation (\ref{eq:poisson}) in terms of excess $\nu$ and Hubble parameter $H$:
%%%%%%
\begin{equation}
\frac{\nabla^2\Phi}{a^2} = 4\pi G\, W \left(\nu-3H\varphi\right) .
\label{eq:poisson1}
\end{equation}
%%%%%%
We see also here that for relativistic fluids the enthalpy density $W$ plays the role of the mass density. Note that we wrote the Poisson equation (\ref{eq:poisson}) or (\ref{eq:poisson1}) assuming only one fluid. In reality, we need to sum over the contributions from all fluids, as gravity is universal, whereas the Bernoulli equation (\ref{eq:bernoulli0}) applies to each fluid separately. 

The same density deviations $\delta\rho+\dot{\rho}\,\varphi$ that drive sound waves also create the gravitational potential. What are they?   Consider the average density $\rho$ in a frame locally moving with the fluid. The density $\rho$ varies with cosmological time $t$, but time is transformed in the moving frame. We obtain from the Lorentz transformation \cite{LL2} for small velocities: $t'\sim t + \bm{v}\cdot \delta\bm{r} \sim t+\varphi$. Consequently, to linear order, $\dot{\rho}\varphi$ is the deviation of the average density from the density comoving with the fluid, and $\delta\rho+\dot{\rho}\,\varphi$ is the total deviation --- due to spatial changes in density and velocity. What counts in both fluid mechanics and gravity is the deviation from the comoving density. Note that the relativistic correction $\dot{\rho}\varphi$ goes with $c^{-2}$ (as it comes from the time transformation) and so does the correction $-3\Phi$ in the equation (\ref{eq:continuity}) of continuity (as $\Phi$ is the Newtonian potential divided by $c^2$ \cite{LL2}). Like in the Newton equation (\ref{eq:newton}) for the universe, the relativistic corrections in both fluid mechanics and gravity are minimal and of second order in $1/c$. The primary role of relativity is less conspicuous but more profound: relativity makes its primary appearance in the $4/3$ factor in the radiation enthalpy $W_\mathrm{R}$  of Eq.~(\ref{eq:radmat}) important in the effective speed of sound, Eq.~(\ref{eq:sound}), and in the very fact that radiation carries weight and causes gravity.

We outline now how the Poisson equation (\ref{eq:poisson}) follows from general relativity and thermodynamics in the regime relevant to cosmology: weak gravity, small fluctuations and non--relativistic velocities. Our starting point are Einstein's field equations \cite{LL2}:
%%%%%%
\begin{equation}
R^\alpha_\beta - \frac{R}{2}\,\delta^\alpha_\beta = 8\pi G\, T^\alpha_\beta 
\label{eq:einstein}
\end{equation}
%%%%%%
where $R^\alpha_\beta$ denotes the Ricci tensor and $R$ the curvature scalar \cite{LL2}. We substitute our expressions for the metric and the energy--momentum tensor into the definitions \cite{LL2} and linearize in the small quantities: the gravitational potential $\Phi$ (for weak gravity), the density deviation $\delta\rho$ (for small fluctuations) and the velocity potential $\varphi$ (for small velocities). In zeroth order, the Einstein equations reduce to the Friedman equation (\ref{eq:f1}) and the thermodynamic relation (\ref{eq:f2w}). In first order, we obtain three equations:
%%%%%%
\begin{gather}
\frac{1}{a^2} \left(\nabla^2\Phi - 3\dot{a}\partial_ta\Phi \right) = 4\pi G \delta\rho \,,\quad
\partial_t a^2 \partial_t a \Phi = 4\pi G a^3 (W\Phi +c_\mathrm{s}^2 \delta\rho) \,,\nonumber\\
\partial_t a \Phi = -4\pi G W a\, \varphi\,.
\label{eq:einrel}
\end{gather}
%%%%%%
The first two equations are easily recognized as the Poisson equation (\ref{eq:poisson}) and the Bernoulli equation (\ref{eq:bernoulli0}) if the third equation is valid. We only need to work out what this strange relationship between the gravitational and the velocity potential actually means. We apply $\nabla^2$ to both sides, use the Poisson equation (\ref{eq:poisson1}) and divide by the prefactor:
%%%%%%
\begin{equation}
\frac{\nabla^2\varphi}{a^2}= \frac{1}{a^3}\, \partial_t\, a^3 W (\nu -3H\varphi) \,.
\end{equation}
%%%%%%
The time derivative of $\varphi$ is given by the Bernoulli equation (\ref{eq:bernoulli}) while for $\partial_t (a^3 W)$ we split $W$ into the density $\rho$ and pressure $W-\rho$, apply the thermodynamic relation (\ref{eq:f2w}) and the definition (\ref{eq:defsound}) of the speed of sound, and get $\partial_t\, a^3 W = -3a^3 c_\mathrm{s}^2 H W$. We thus obtain in total:
%%%%%%
\begin{equation}
\frac{\nabla^2\varphi}{a^2}= \dot{\nu} - 3\dot{H}\varphi +3H\Phi\,.
\label{eq:eincont}
\end{equation}
%%%%%%
Finally, we turn to the third of Eqs.~(\ref{eq:einrel}) again and write for the left--hand side $a^{-1}\partial_t a \Phi=\dot{\Phi} + H\Phi$ and for the right--hand side $4\pi G W \varphi=-\dot{H}\varphi$. Here we have used the thermodynamic relation (\ref{eq:f2w}) and differentiated the Friedman equation (\ref{eq:f1}). Consequently, Eq.~(\ref{eq:eincont}) gives just the equation of continuity (\ref{eq:continuity}). The Einstein equations (\ref{eq:einstein}) of gravity all reduce to the Poisson equation (\ref{eq:poisson}). 

\subsection{Light and matter}

We have deduced the equations of motion for sound waves of light and matter. These waves imprint their traces in the Cosmic Microwave Background (CMB) as temperature fluctuations. The fluctuations $\delta T_0/T_0$ we see today are the same as the original $\delta T/T$ in the light emitted at the time of last scattering, because temperature is inversely proportional to scale factor, Eq.~(\ref{eq:temperature}). Two terms contribute mainly to $\delta T/T$: the intrinsic temperature variation $\delta T_\mathrm{int}/T$ due to the density deviation $\delta\rho$ and the temperature variation $\delta T_\mathrm{grav}/T$ due to the gravity $\Phi$ (called the Sachs--Wolfe effect \cite{SachsWolfe}). The temperature varies intrinsically with density, because according to the Stefan--Boltzmann law (\ref{eq:stefanboltzmann}) the density depends on temperature and vice versa: $T$ goes with $\rho^{1/4}$ for radiation. Hence $\delta T_\mathrm{int}/T=\delta\rho/4$ or, using the thermodynamic relation (\ref{eq:relation}) and Eq.~(\ref{eq:radmat}), we have $\delta T_\mathrm{int}/T=\nu/3$. This is the temperature variation of the emitted light. The light also gets gravitationally redshifted by the local variation of the gravitational field. Gravity changes the local measure of time: $\sqrt{g_{00}}\sim1+\Phi$ according to the space--time metric (\ref{eq:metric1}) for weak $\Phi$. The gravitational redshift in the emission frequency appears as a blueshift in the received temperature in the Boltzmann factors $\exp(-\hbar\omega/k_\mathrm{B}T)$. We thus have $\delta T_\mathrm{grav}/T=\Phi$ and get in total:
%%%%%%
\begin{equation}
\frac{\delta T}{T} = \frac{\nu}{3} + \Phi \,.
\label{eq:deltaT}
\end{equation}
%%%%%%

The CMB is formed in an expansion era dominated by radiation and matter where we can safely ignore the cosmological constant. Here it is wise to represent the scale factor $a$ and cosmological time $t$ as $n$ and $\vartheta$ according to the scaling law~(\ref{eq:rmscaling}). We also scale the comoving coordinates $\bm{r}$ the same way as time. Velocities are not changed, but the velocity potential gets scaled like time. We thus introduce the scaled velocity potential $\psi$ as
%%%%%%
\begin{equation}
\psi = \frac{\Omega_\mathrm{M}^2}{\Omega_\mathrm{R}^{3/2}}\,H_0\, \varphi \,.
\end{equation}
%%%%%% 
The gravitational potential $\Phi$ carries the physical dimensions of a velocity squared and thus remains unchanged.  Inspecting the dynamical equations (\ref{eq:continuity}) and (\ref{eq:bernoulli}) we see that they are invariant, if the Hubble parameter $H$ is also scaled (like inverse time). The scaled Hubble parameter reads (using the same symbol for simplicity):
%%%%%%
\begin{equation}
H = n^{-2}\sqrt{1+n}\,.
\label{eq:hubblescaled}
\end{equation}
%%%%%% 
In the Poisson equation (\ref{eq:poisson1}) we apply Eq.~(\ref{eq:radmat}) for relating the enthalpy densities $W_m$ to the mass densities $\rho_m$ and express $(4\pi/3)G\rho_m$ by $\Omega_\mathrm{R} H_0^2/ a^4$ and $\Omega_\mathrm{M} H_0^2/ a^3$ according to the cosmological standard model (Sec.~3). The left--hand side of the Poisson equation (\ref{eq:poisson1}) then absorbs all factors of $\Omega_\mathrm{R}$, $\Omega_\mathrm{M}$ and $H_0$ in the scaled units of Eq.~(\ref{eq:rmscaling}). Consequently, for the scaled quantities, both fluid dynamics and gravity are independent on the actual cosmic parameters. The parameters only appear in the scaling (\ref{eq:rmscaling}) relating the physical scale factor $a$ to $n$ and the cosmological time $t$ to $\vartheta$.

The universe is expanding with time, but on average space is homogeneous and flat. We can thus use spatial Fourier  transformation to decompose the fluctuations into plane waves with wavenumber $k$. For the Fourier--transformed quantities we use the same notation as for the original ones. We scale $k$ like an inverse time and obtain for the scaled quantities from the continuity equation (\ref{eq:continuity}) and the Bernoulli equation (\ref{eq:bernoulli}):
%%%%%%
\begin{equation}
\partial_\vartheta (\nu-3\Phi) = \frac{k^2\psi}{n^2} \,,\quad
\partial_\vartheta \psi + \Phi + c_\mathrm{s}^2(\nu -3H\psi) = 0 \,.
\label{eq:motionscaled}
\end{equation}
%%%%%%
Cosmologists measure time in terms of redshift, because time evolution and cosmic expansion are related. We may thus as well write the equations of motion (\ref{eq:motionscaled}) in terms of the scale factor $n$:
%%%%%%
\begin{equation}
\partial_\vartheta = n H\,\partial_n = n^{-1}\sqrt{1+n} \,\partial_n \,.
\label{eq:evolution}
\end{equation}
%%%%%%
Each cosmic fluid obeys a separate set of equations of motion. We have three such fluids: light coupled to baryonic matter, neutrino radiation and dark matter. The fraction of light in all radiation is given by $1/\eta_\mathrm{R}$ where $\eta_\mathrm{R}$ denotes the prefactor in the Stefan--Boltzmann law (\ref{eq:stefanboltzmann}). Baryonic matter constitutes the fraction $\eta_\mathrm{B}$ of all matter with $\eta_\mathrm{B}=0.156$ according to the CMB measurements \cite{CMBPlanck}. In the speed of sound of the light--matter fluid, Eq.~(\ref{eq:sound}), only light and baryonic matter should enter with common factor
%%%%%%
\begin{equation}
\eta = \eta_\mathrm{R} \,\eta_\mathrm{B} \,.
\label{eq:eta}
\end{equation}
%%%%%%
For pure radiation $c_\mathrm{R}=1/\sqrt{3}$ in our scaled units. We thus obtain from Eqs.~(\ref{eq:radmat}) and $\rho_\mathrm{M}/\rho_\mathrm{R}=n$ the speed of sound:
%%%%%%
\begin{equation}
c_\mathrm{s} = \frac{1}{\sqrt{3\left(1+\frac{3\eta}{4}\,n\right)}} \,.
\label{eq:cmbsound}
\end{equation}
%%%%%%
This is the speed of sound that shapes the undulating correlations of the CMB fluctuations (Fig.~\ref{fig:correlations}). The phase of the waves we denote by $\varphi$ (not to be confused with the velocity potential). The phase is given by the integral:
%%%%%%
\begin{equation}
\varphi= k \int_0 c_\mathrm{s}\,\mathrm{d}\tau = \frac{4k}{3\sqrt{\eta}}\,\ln \frac{\sqrt{1+\frac{3\eta}{4}\,n}+\sqrt{\frac{3\eta}{4}(1+n)}}{1+\sqrt{\frac{3\eta}{4}}} \,.
\label{eq:phasen}
\end{equation}
%%%%%% 
The phase we see in the undulations of the CMB spectrum is the phase at the time of last scattering:
%%%%%%
\begin{equation}
\varphi_\mathrm{*} = \varphi \quad \mbox{for} \quad n=n_\mathrm{*} \,.
\label{eq:phaseexpression}
\end{equation}
%%%%%% 
Expression (\ref{eq:phaseexpression}) depends on $n_\mathrm{*}=(\Omega_\mathrm{R}/\Omega_\mathrm{M})a_\mathrm{*}$ where $a_\mathrm{*}$ relates the measured temperature $T_0$ of the CMB to the temperature $T_\mathrm{*}$ of last scattering one obtains from theory \cite{Weinberg}. As $a_\mathrm{*}=T_0/T_\mathrm{*}$ we have:
%%%%%%
\begin{equation}
n_\mathrm{*} = \frac{\Omega_\mathrm{M}}{\Omega_\mathrm{R}}\,\frac{T_0}{T_\mathrm{*}} \,.
\end{equation}
%%%%%% 
Expression (\ref{eq:phaseexpression}) also depends via formula (\ref{eq:phasen}) and relation (\ref{eq:eta}) on the fraction $\eta_\mathrm{B}$ of the baryonic matter and therefore on the dark--matter content $1-\eta_\mathrm{B}$. By fitting $\Omega_\mathrm{R}/\Omega_\mathrm{M}$ and $\eta_\mathrm{B}$ to the correlation curve governed by the phase (\ref{eq:phaseexpression}) one could retrieve the cosmic parameters. 

Note however that the other fluids, neutrino radiation N and dark matter D, also actively contribute to the dynamics, not only passively through the fraction $\eta$, because they interact with the light--matter waves by gravity. Given a gravitational potential $\Phi$, each fluid has its own separate dynamics, described by  Eqs.~(\ref{eq:motionscaled}) and (\ref{eq:cmbsound}) and 
%%%%%%
\begin{eqnarray}
\partial_\vartheta (\nu_\mathrm{N}-3\Phi) = \frac{k^2\psi_\mathrm{N}}{n^2} \,&,&\quad
\partial_\vartheta \psi_\mathrm{N} + \Phi +\frac{1}{3}(\nu_\mathrm{N} -3H\psi_\mathrm{N}) = 0 \,,\\
\partial_\vartheta (\nu_\mathrm{D}-3\Phi) = \frac{k^2\psi_\mathrm{D}}{n^2} \,&,&\quad
\partial_\vartheta \psi_\mathrm{D} + \Phi = 0 \,,
\end{eqnarray}
%%%%%%
but all fluids contribute to the overall gravitational field:
%%%%%%
\begin{equation}
-\frac{2k^2}{3}\, \Phi = \left(\frac{4\eta_\gamma}{3n} + \eta_\mathrm{B}\right)\frac{\nu-3H\psi}{n} + \frac{4\eta_\mathrm{N}} {3}\,\frac{\nu_\mathrm{N}-3H\psi_\mathrm{N}}{n^2} +\eta_\mathrm{D}\frac{\nu_\mathrm{D}-3H\psi_\mathrm{D}}{n}
\label{eq:gravity}
\end{equation}
%%%%%%
according to their relative weights $\eta_m$ with $\eta_\gamma=1/\eta_\mathrm{R}$, $\eta_N=1-\eta_\gamma$ and $\eta_\mathrm{D}=1-\eta_\mathrm{B}$. While light--matter waves propagate with the effective $c_\mathrm{s}$ of Eq.~(\ref{eq:cmbsound}), neutrino perturbations propagate with speed $1/\sqrt{3}$ and dark matter follows inertia. Neutrinos and dark matter are getting out of sync with light and baryonic matter and perturb the light--matter dynamics by their gravity. The full fluid--mechanical theory (Fig.~\ref{fig:waves}) describes the phase of the CMB well, but not its amplitude, in particular for large spherical harmonics beyond the first peak in the spectrum (Fig.~\ref{fig:correlations}) that correspond to large $k$. The reason is that that the time of last scattering is not perfectly sharp. The initially opaque plasma does not become transparent in an instant, but over a characteristic time (the Silk time \cite{Silk}) that depends on the recombination dynamics. As a result the correlations in the CMB are not perfectly coherent, a mechanism called Silk damping \cite{Silk} that washes out the undulations in the correlation spectrum (Fig.~\ref{fig:correlations}). An accurate comparison between theory and observation --- sufficiently accurate for inferring the cosmic parameters (\ref{eq:parameters}) --- is possible (Fig.~\ref{fig:correlations}) but requires a finer level of resolution than fluid mechanics. The particles need to be sorted according to their velocities and energies and their interaction described by relativistic Boltzmann equations \cite{Dodelson}.

\subsection{Solutions}

Nevertheless, with fluid mechanics we can deduce an exact statement about fluctuations in the radiation--dominated era ($n\ll 1$) that is important for defining the initial conditions. We will also derive an approximate solution for arbitrary $n$ and $k\gg 1$ that reveals not only the phase, but also the amplitude structure. Silk damping \cite{Silk} can be included \cite{Weinberg} by considering a statistical ensemble of times of last scattering (corresponding to an ensemble of $n_\mathrm{*}$). Consider the radiation--dominated period. Inflation is believed to have amplified vacuum fluctuations such that they become macroscopic. The unstructured noise of the amplified vacuum is uniform for all particles such that we can assume \cite{Weinberg}:
%%%%%%
\begin{equation}
\nu = \nu_\mathrm{N} = \nu_\mathrm{D} \,.
\label{eq:initial}
\end{equation}
%%%%%%
The noise has a scale--invariant power--law spectrum \cite{Weinberg}. We multiply this spectrum with a solution of the dynamics of uniform initial amplitude for all Fourier components. We thus assume:
%%%%%%
\begin{equation}
\nu(0) =\nu_0 \quad\mbox{for all $k$.}
\label{eq:initial0}
\end{equation}
%%%%%%
Now, for $n\ll 1$ we can ignore all matter contributions. The radiation contributions have been the same initially, and as light has not interacted with matter yet, and neutrinos never do, they remain the same. We thus get from (\ref{eq:gravity}) for the gravitational potential: 
%%%%%%
\begin{equation}
-\frac{k^2}{2}\, \Phi = \frac{\nu-3H\psi}{n^2} \,.
\label{eq:phi0}
\end{equation}
%%%%%%
We include the gravity--induced change in spatial density in the excess $\nu$ by defining the total excess 
%%%%%%
\begin{equation}
\mu = \nu - 3\Phi
\label{eq:mu}
\end{equation}
%%%%%%
and solve Eq.~(\ref{eq:phi0}) as a function of $\mu$ for $\Phi$:
%%%%%%
\begin{equation}
\Phi = -\frac{2(\mu-3 H \psi)}{6+k^2 n} \,.
\label{eq:phi1}
\end{equation}
%%%%%%
One verifies that 
%%%%%%
\begin{equation}
\mu = -3\nu_0\left(\cos\varphi- \frac{2\sqrt{3}}{kn}\,\sin\varphi\right) 
\label{eq:muR}
\end{equation}
%%%%%%
with phase
%%%%%%
\begin{equation}
\varphi = \frac{k n}{\sqrt{3}}
\label{eq:phaseR}
\end{equation}
%%%%%%
and velocity potential 
%%%%%%
\begin{equation}
\psi = \frac{n^3 H}{k^2}\,\partial_n \mu 
\label{eq:psiR}
\end{equation}
%%%%%%
is an exact solution of the Bernoulli equation
%%%%%%
\begin{equation}
n H \partial_n \psi + \Phi +\frac{1}{3}(\nu -3H\psi) = 0
\end{equation}
%%%%%%
for the gravitational potential of Eq.~(\ref{eq:phi1}) and the Hubble parameter $H=n^{-2}$. In fact, it is the only regular solution that does not diverge for $n\rightarrow 0$. The other, divergent solution of the fundamental system we get if we replace $\varphi$ by $\varphi+\pi/2$. For the convergent solution, $\mu(0)= 3\nu_0$ and $H\psi\rightarrow \nu_0/3$ with $\psi$ defined in Eq.~(\ref{eq:psiR}). From Eq.~(\ref{eq:phi1}) follows $\Phi\rightarrow -2\nu_0/3$ and hence $\nu=\mu+3\Phi\rightarrow \nu_0$. We see that $\nu$ satisfies the initial condition (\ref{eq:initial0}). For the initial temperature variation (\ref{eq:deltaT}) we get $\delta T/T\rightarrow -\nu_0/3$. Condition (\ref{eq:initial}) then implies that all cosmic fluids move initially with equal velocity potential 
%%%%%%
\begin{equation}
\psi = \psi_\mathrm{N} = \psi_\mathrm{D} \quad\mbox{and}\quad \psi\sim\frac{\nu_0}{3}\,n^2 \,.
\label{eq:initialpsi}
\end{equation}
%%%%%%
We have thus determined the initial conditions after inflation without any assumptions made, apart from regularity and equal particle number across all particle species for each Fourier component $k$. 

One can also obtain approximate solutions of the fluid--mechanical equations for large wavenumbers ($k\gg 1$) that naturally connect to the exact initial solution for all $k$. Let us briefly outline how this is done. Consider the simplest case, a hypothetical universe without neutrinos and without dark matter. In this case, light and matter form a closed system not perturbed from other components. We have for the gravitational potential: 
%%%%%%
\begin{equation}
-\frac{2k^2}{3}\, \Phi = \left(\frac{4}{3n} + 1 \right)\frac{\nu-3H\psi}{n} \,.
\label{eq:gravityrm}
\end{equation}
%%%%%%
As before, we focus on the total excess $\mu$ defined in Eq.~(\ref{eq:mu}) and solve Eq.~(\ref{eq:gravityrm}) for $\Phi$ as a function of $\mu$. Inspired by our exact solution (\ref{eq:muR}) for the radiation--dominated era, we make the ansatz
%%%%%%
\begin{equation}
\mu = -3\nu_0\, {\cal A}\left(\cos\varphi - \frac{{\cal B}}{k}\,\sin\varphi\right) 
\label{eq:solution}
\end{equation}
%%%%%%
with $\varphi$ given by Eq.~(\ref{eq:phasen}) for $\eta=1$,
%%%%%%
\begin{equation}
\varphi= \frac{4k}{3}\left(\mathrm{arsinh}\sqrt{3\,(1+n)}-\mathrm{arsinh}\sqrt{3} \right) ,
\label{eq:phasen1}
\end{equation}
%%%%%%
and ${\cal A}$ and ${\cal B}$ to be determined. For the sake of our calculation we supplement $\mu$ by an imaginary part, and write
%%%%%%
\begin{equation}
\mu = -3\nu_0\, {\cal A}\, (1- \mathrm{i}{\cal B}/k) \,\mathrm{e}^{-\mathrm{i}\varphi} \,.
\label{eq:solutioncomplex}
\end{equation}
%%%%%%
The velocity potential $\psi$ we get from Eq.~(\ref{eq:psiR}). We then insert $\Phi$ and $\psi$ in the Bernoulli equation 
%%%%%%
\begin{equation}
n H \partial_n \psi + \Phi + c_\mathrm{s}^2(\mu + 3\Phi -3H\psi) = 0 
\label{eq:bernoullin}
\end{equation}
%%%%%%
with Hubble parameter (\ref{eq:hubblescaled}) and speed of sound (\ref{eq:cmbsound}) with $\eta=1$. We extract the overall phase factor $\exp(-\mathrm{i}\varphi)$ and expand in a power series in $k$ we solve term by term, starting with the highest. The choice (\ref{eq:phasen1}) for the phase of the wave takes care of the dominant term. The next term determines the amplitude: 
%%%%%%
\begin{equation}
{\cal A} = \left(1+\frac{3n}{4}\right)^{-1/4} 
\label{eq:amp}
\end{equation}
%%%%%%
that approaches unity for $n\rightarrow 0$. Finally we go one term lower and calculate ${\cal B}$. We get 
%%%%%%
\begin{equation}
\frac{\mathrm{d}{\cal B}}{\mathrm{d}n} = -\frac{\sqrt{3}}{64}\,\frac{1024+3n(640+389n+75n^2)}{n^2(1+n)^{1/2}(4+3n)^{3/2}}
\end{equation}
%%%%%%
and obtain after integration:
%%%%%%
\begin{equation}
{\cal B} = {\cal B}_0 + {\cal B}_1 + {\cal B}_2
\end{equation}
%%%%%%
with the contributions
%%%%%%
\begin{eqnarray}
{\cal B}_0 &=& \frac{\sqrt{3}}{32}\left(93+\frac{128}{n}\right)\sqrt{\frac{1+n}{4+3n}} ,
\nonumber\\
{\cal B}_1 &=& \frac{\sqrt{3}}{2} \,\ln \frac{8+7n+4\sqrt{1+n}\,\sqrt{4+3n}}{n} \,,
\nonumber\\
{\cal B}_2 &=& - \frac{75}{64}\,\ln\left(7+6n+2\sqrt{1+n}\,\sqrt{12+9n}\right) .
\end{eqnarray}
%%%%%%
The details are not important here, but we see that ${\cal B}$ matches the exact solution (\ref{eq:psiR}) for $n\ll 1$; there ${\cal B}_0$ dominates and approaches exactly the corresponding term in Eq.~(\ref{eq:psiR}). For $n\gg 1$ the terms ${\cal B}_0$ and ${\cal B}_1$ become constants, while ${\cal B}_2 \sim-\frac{75}{64}\ln(12n)$. The phase $\varphi$ and amplitude ${\cal A}$ of Eqs.~(\ref{eq:phasen1}) and (\ref{eq:amp}) also approach the phase and amplitude of the initial wave of Eq.~(\ref{eq:muR}) for $n\ll 1$. Consequently, the solution (\ref{eq:solution}) naturally matches the initial conditions.

Our solution describes the asymptotics of the light--matter dynamics  for large wave\-numbers --- in the hypothetical universe entirely dominated by light and baryonic matter. Neutrinos and dark matter are going to perturb it. Nevertheless, the solution allows us to distinguish the three different physical mechanisms that shape the undulations in the CMB. The most prominent feature is the phase (``phase rules the waves''). The phase $\varphi$ is given by the wavenumber $k$ times the speed of sound $c_\mathrm{s}$ integrated over conformal time $\tau$.  As $\varphi$ grows linearly with $k$ there is no spatial dispersion. The phase just contains the accumulated effect of the variation in $c_\mathrm{s}$ due to the expanding universe and of the expansion itself. Initially, the cosmic sound waves are entirely made of radiation with constant sound speed $1/\sqrt{3}$ and $n\sim\tau$, such that $\varphi$ grows linearly in $n$. For $n\gg 1$ matter dominates, $c_\mathrm{s}$ of Eq.~(\ref{eq:cmbsound}) falls like $n^{1/2}$ and $n$ of Eq.~(\ref{eq:nrmtau}) goes like $\tau^2$, which produces only logarithmic growth of $\varphi$. 

Now turn to the amplitude, Eq.~(\ref{eq:amp}). Initially the amplitude is constant, as Eq.~(\ref{eq:muR}) shows, and then ${\cal A}$ decreases modestly as $n^{-1/4}$. The variation of the amplitude is due to the variation of the speed of sound $c_\mathrm{s}$ with expansion $n$. The initial amplitude is constant, because in the radiation--dominated era light co--propagates with the expanding universe, $n\sim\tau$. Not only the average radiation co--propagates, but perturbations, too. They would evolve as $\nu=\nu_0\cos(\bm{k}\cdot\bm{r}-kc_\mathrm{s}\tau)$ if $c_\mathrm{s}=1/\sqrt{3}$ as one verifies in the equations of motion without gravity. The cosmic sound would consist of perfect plane waves, all synchronized, were it not for gravity. 

Gravity generates the second term $\propto (kn)^{-1} \sin\varphi$ in the exact initial solution (\ref{eq:muR}) and the ${\cal B}$--term in formula (\ref{eq:solution}). We see from the complex form (\ref{eq:solutioncomplex}) that for $k\gg 1$ the amplitude of the wave is not much affected, but the phase gets an additional contribution: $-{\cal B}/k$. Instead of growing linearly with $k$ this phase shift goes like $k^{-1}$. One can easily understand why gravity induces such a $k^{-1}$ shift: According to the Poisson equation (\ref{eq:gravityrm}) $\Phi$ falls like $k^{-2}$. The gravitational potential causes a perturbation competing with the energy $-nH\partial_n\psi$ in the Bernoulli equation (\ref{eq:bernoullin}).  The energy goes with $k^{-1}$ as $\psi$ goes with $k^{-2}$ according to the equation (\ref{eq:psiR}) of continuity. The result is the phase shift $-{\cal B}/k$. Gravity delays acoustic waves, shifting their phases, long waves with low $k$ stronger than short ones with high $k$. Gravity de--synchronizes the cosmic sound. 

These features, although not sufficiently accurate for fitting the cosmic parameters to the correlations of the CMB, give intuition and insight into an otherwise opaque subject confined to the black box of the computer. It is utterly remarkable how, in cosmology, relativistic fluid mechanics and general relativity are reducible to good--old, hands--on physics without compromise in intellectual accuracy and depth. ``The most incomprehensible thing about the universe is that it is comprehensible'' (Einstein).

\end{document}